\documentclass[preprint,pteplogo]{ptephy_v2}

\preprintnumber{XXXX-XXXX} 

\usepackage[margin=1in]{geometry}
\usepackage{booktabs}
\usepackage{enumitem}
\usepackage{amsmath,amsthm}
\usepackage{hyperref}
\hypersetup{colorlinks=true,linkcolor=blue,citecolor=blue,urlcolor=blue}
\usepackage{bm}
\usepackage{graphicx}
\graphicspath{{figures/}}
\usepackage[dvipsnames]{xcolor}
\usepackage{mathrsfs}
\newtheorem{proposition}{Proposition}
\theoremstyle{remark}
\newtheorem{remark}{Remark}

\newcommand{\half}{\tfrac12}
\newcommand{\be}[1]{\begin{equation} \label{#1}}
\newcommand{\ee}{\end{equation}}
\newcommand{\bea}{\begin{eqnarray}}
\newcommand{\bean}{\begin{eqnarray*}}
\newcommand{\eea}{\end{eqnarray}}
\newcommand{\eean}{\end{eqnarray*}}
\newcommand{\ba}{\begin{array}}
\newcommand{\ea}{\end{array}}
\newcommand{\bel}{\begin{align}}
\newcommand{\eel}{\end{align}}

\begin{document}

\title{Finite-time geometric focusing and current-sheet precursors in helical magnetic flux ropes}


\author{Hyeong-Chan Kim}
\affil{School of Liberal Arts and Sciences, Korea National University of Transportation, Chungju 27469, Korea \email{hckim@ut.ac.kr}}




\begin{abstract}%
\noindent
We study a cylindrical helical magnetic flux rope subject to the requirement that its electromagnetic field retain the same local two-plane structure as it evolves. 
On the nondegenerate branch considered here, \(F\wedge F\neq0\), equivalently \(\mathbf E\cdot\mathbf B\neq0\), the Maxwell tensor splits canonically into two decomposable parts spanning mutually orthogonal two-planes; we require the magnetic-type part to keep a fixed radial leg and a helical direction, with a prescribed longitudinal profile. 
Imposing the source-free Maxwell equations $dF=0$ on the total field then yields a convex scalar conservation law for the field-line pitch variable $\Phi=r\,|B_{\hat\phi}|/B_{\hat\chi}$, whose ratio to $r$ is the tangent of the local field-line pitch angle. 
Smooth compressive pitch profiles therefore steepen, and the radial characteristic map loses invertibility after a finite accumulated drive while the pitch and the field amplitudes can remain finite. 
Provided an explicitly computable prefactor does not vanish, the Amp\`ere current required to support the rope diverges at that point. 
We call this endpoint a current-sheet precursor, because the analysis does not show that a self-consistent plasma reaches it, nor which layer resistive, Hall, or kinetic physics would select. 
The focusing survives the addition of radial magnetic through-flux. 
An exact homologous family shows that pitch-gradient blow-up is not the essential feature: the characteristic Jacobian vanishes and the rotation rate of the unit helical direction diverges while $\partial_r \Phi \to 0$. 
Since  \(\mathbf E\!\cdot\!\mathbf B\neq0\), no four-velocity satisfies $F_{ab} u^b =0$ and no frame removes the electric field -- an algebraic property that fixes no resistivity, dissipation rate, or reconnection model.
\end{abstract}

\subjectindex{Electromagnetism, Plasma astrophysics, Nonlinear phenomena(self-organizations, chaos, turbulences)}

\maketitle

\section{Introduction}
\label{sec:intro}

Magnetic flux ropes are coherent structures in which longitudinal and
azimuthal magnetic fields combine into a radially organized helical
configuration.  Their stability and breakdown are central problems in
laboratory, space, and astrophysical plasmas, and are commonly discussed in
terms of ideal normal-mode instabilities, loss of force balance, current-sheet
formation, magnetic reconnection, and dissipative transport
\cite{Lundquist1951,GoldHoyle1960,Newcomb1958,PriestForbes2000,Freidberg2014}.
In the solar flux-rope context these two routes are usually contrasted as
loss of equilibrium, or \emph{catastrophe}, on the one hand
\cite{ForbesIsenberg1991} and ideal instability---helical kink or
torus---on the other \cite{TorokKliem2005,KliemTorok2006}; the two criteria
have been compared directly and can coincide in some
configurations~\cite{DemoulinAulanier2010,KliemLinForbesPriest2014}.
The word \emph{catastrophe} will be used here in neither of these senses,
but in the PDE sense of a finite-time loss of smoothness.

We address a different question. Suppose that a cylindrical helical magnetic structure is required to preserve its local field-plane organization during the evolution. Can this geometric organization itself develop a finite-time singularity, even when the electromagnetic field amplitudes remain finite? In particular, can neighboring radial characteristics focus to a caustic and thereby concentrate the electric current required to support the helical structure?

We show that the answer is affirmative within the restricted helical branch
studied here.  For a fixed longitudinal magnetic flux, the field-line
pitch $\Phi$ obeys a nonlinear first-order transport equation whose characteristic speed depends on the pitch itself.  Smooth compressive profiles can therefore
develop a finite-time characteristic caustic: neighboring radial labels
overtake one another and the radial characteristic map loses local
invertibility.  The pitch and the electromagnetic field amplitudes can remain
finite at the first crossing, while the spatial rotation rate of the
normalized helical field plane becomes singular.  Under the additional
nondegeneracy condition that the coefficient multiplying the singular pitch
gradient in the Amp\`ere current does not vanish, the physical current also
diverges.  The formal endpoint is therefore interpreted as a
\emph{current-sheet precursor}.
 Whether a self-consistent plasma actually
reaches this caustic depends on the material response.  Resistive, Hall,
kinetic, dispersive, or other constitutive effects may modify the magnetic-flux
evolution or reorganize the electromagnetic structure before characteristic
crossing occurs.  Our result is therefore conditional on the persistence of
the helical branch up to the first collapse: under this assumption, the
required current generically becomes singular while the electromagnetic field
itself may remain algebraically regular.

The focusing is also not peculiar to the simplest purely spatial ansatz.
Allowing a temporal component in the helical sector introduces radial magnetic
through-flux in the orthogonal field plane while preserving the fixed radial
leg.  The longitudinally symmetric (\(\chi\)-independent) through-flux leaves
the original radial characteristic transport unchanged, and more general dependence provides flux redistribution
without introducing an intrinsic smoothing scale.  The focusing mechanism is
therefore robust within this natural fixed-radial-leg flux-rope class.
Allowing the longitudinal flux itself to evolve gives a more substantial
escape channel because it changes the characteristic map.  Nevertheless, an
exact homologous family still reaches a global caustic: the characteristic
Jacobian vanishes while the Eulerian pitch slope tends to zero and the
normalized field-plane rotation rate diverges.  
Thus, pitch-gradient blow-up is only one realization of the collapse.  Its defining kinematic feature is instead the finite-time \emph{loss of invertibility} of the characteristic map,
while the divergent rotation of the normalized helical field plane provides
its intrinsic geometric signature within the regular helical branch.

We now describe the geometric constraint that produces this dynamics.
A generic non-null electromagnetic field can locally be resolved into two
simple contributions associated with two metric-orthogonal two-planes
\cite{MisnerWheeler1957,Stazi2006}.  We denote these sectors by
\(G\) and \(H\), with the total field satisfying the Bianchi identity,
\begin{equation}
        F=G+H , \qquad dF=0  \quad \Longleftrightarrow \quad  dG=-dH ,
        \label{closure}
\end{equation}
so that the exterior derivatives of $G$ and $H$ cancel each other.
Here \emph{simple} means decomposable, $G=\alpha\wedge\beta$, so that each
sector spans a single two-plane.
The helical flux-rope structure is assigned to \(G\), while \(H\) occupies
the complementary metric-orthogonal plane. 

For a non-null $F$ the orthogonal pair $(G,H)$ is unique-- it is the canonical (Misner--Wheeler) form. The restriction imposed here is not orthogonality, which is automatic, but that the magnetic-type blade retain the radial-helical form, with a fixed radial leg, at every stage of the evolution.
It is useful to fix the terminology once. Existence and uniqueness of the
canonical pair require only that $F$ be non-null; the further condition
$F\wedge F\neq0$, used throughout as the definition of the
\emph{nondegenerate} branch, states in addition that \emph{both} sectors are
nonzero, so that $H$ does not collapse and the two-plane pair is genuinely
complementary. A logically separate nondegeneracy condition,
$\Delta:=\half G_{ab}G^{ab}\neq0$, refers to the sector $G$ alone and is
introduced only in Sec.~\ref{sec:radial-extension}.
This requirement is stronger than merely asking for the existence of some
closed electromagnetic continuation.  
A more general continuation may remain regular by allowing additional deformations of the local field planes beyond the flux-rope structure considered here.  
Such a continuation represents a structural reorganization of the electromagnetic field rather than persistence of the same radial--helical branch.  
More general local reconstructions of this kind are possible for compatible data~\cite{Kim2}.

For the restricted branch considered here, metric orthogonality fixes the
relative components of the second field contribution algebraically.  
When this relation is substituted into \(dF=0\), the Bianchi identity becomes a
nonlinear transport equation for the helical pitch.  
Thus the nonlinearity is not attributed to the Bianchi identity alone.  
It arises from imposing Bianchi closure while requiring the radial--helical structural class of \(G\) and its orthogonal pairing with \(H\) to persist.  
The geometric quantity used below to diagnose the resulting loss of smoothness is the rotation rate of the normalized helical plane.  
In Sec.~\ref{sec:helical-fluxtube} this quantity is expressed as a normalized Frobenius-anholonomy scalar, but its physical meaning is simply the local rate at which the unit helical direction rotates.

The same construction also fixes an important algebraic property of the
electromagnetic field.  On the nondegenerate branch considered here, both
simple contributions are nonzero and occupy complementary planes, so that
$ F\wedge F = 2G\wedge H  \neq0 ~\Leftrightarrow~ \mathbf E\cdot\mathbf B\neq0 .$
The electromagnetic tensor therefore has no kernel: there is no timelike
four-velocity \(u^a\) satisfying $ F_{ab}u^b=0 ,$
and hence no local frame in which the electric field vanishes entirely.
In this precise algebraic sense the branch does not satisfy the ideal-MHD
condition.  This statement alone does \emph{not} specify a resistivity,
dissipation rate, entropy production, or reconnection mechanism; those
properties require an additional material constitutive model.

Matter therefore enters at a logically separate stage.  
The geometric construction determines the electromagnetic characteristics and the current required to support the prescribed helical branch, but it does not determine the matter flow or the evolution law of the longitudinal magnetic flux.
Constitutive transport may reshape that flux and prevent the caustic from
being reached.  
Conversely, if matter is assumed to follow the geometric characteristics, the exact homologous solution provides a stringent consistency test: the causal perfect-fluid model examined in this work cannot follow the formal focusing branch all the way to its endpoint.  
The geometric analysis thus identifies a mechanism for loss of the restricted helical organization, while matter determines whether that branch is avoided,
regularized, or replaced by another electromagnetic structure.

\paragraph{Assumptions}-- The analysis rests on the following. 
(i) Flat spacetime in cylindrical coordinates with $U(1)_\phi$ symmetry. 
(ii) The total field $F$ is non-degenerate, $F\wedge F\neq 0$, on the domain considered, so that its canonical decomposition into two orthogonal decomposable sectors $F =G+H$ exists and is unique. 
(iii) The sector $G$ keeps the fixed-radial-leg helical form 
$G=d r\wedge (B d\phi + A d\chi)$ throughout the evolution in Secs.~\ref{sec:helical-fluxtube} and \ref{sec:time-dependent-B}; this is the restriction whose breakdown we study. 
Section~\ref{sec:radial-extension} relaxes this assumption to include a radial electric field. 
(iv) In Sec.~\ref{sec:helical-fluxtube} the longitudinal profile $B = B(r)$ is prescribed and time independent; Sec.~\ref{sec:time-dependent-B} relaxes this. 
(v) No constitutive matter model is imposed in the geometric evolution of Secs.~\ref{sec:helical-fluxtube}--\ref{sec:time-dependent-B}; specific material closures are considered separately in Sec.~\ref{sec:matter-closure} as consistency tests.

The paper is organized as follows.
Section~\ref{sec:helical-fluxtube} derives the nonlinear pitch transport,
the fixed-flux characteristic caustic, its geometric diagnostic, and the
associated current-sheet precursor.
Section~\ref{sec:radial-extension} introduces the maximal simple extension
that preserves the fixed radial leg and examines the effect of radial magnetic
through-flux.
Section~\ref{sec:time-dependent-B} allows the longitudinal flux to evolve,
derives the characteristic and Jacobi equations, and presents the exact
homologous benchmark.
Section~\ref{sec:matter-closure} discusses the current-support and material
consistency problem.
Section~\ref{sec:discussion} summarizes the physical interpretation and
limitations.

\paragraph{Conventions}---
We use signature \( (-+++) \) and cylindrical coordinates
\(x^a=(t,r,\phi,\chi)\) with metric
\begin{equation}
        ds^2
        =
        -dt^2+dr^2+r^2d\phi^2+d\chi^2 .
\end{equation}
The orientation is chosen so that
$dt\wedge dr\wedge d\phi\wedge d\chi>0$, equivalently
$\epsilon_{tr\phi\chi}=+r$, and $\star$ denotes the Hodge dual defined by this metric and orientation with $\star \mbox{vol}=-1,~\star^2=-1$ on two-forms.  We use Heaviside--Lorentz units with $c=1$ and
$4\pi=1$, so the inhomogeneous Maxwell equation is
$J^\mu=\nabla_\nu F^{\nu\mu}$.  The angle $\phi$ is dimensionless while
$t,r,\chi$ carry dimensions of length. 

\section{Nonlinear focusing of a helical magnetic flux rope}
\label{sec:helical-fluxtube}

We begin with the simplest cylindrical branch in which the longitudinal
magnetic profile is fixed while the helical orientation is allowed to
evolve.  The requirement that the helical simple sector remain
metric-orthogonal to its complementary simple sector, together with the
Bianchi identity for the total electromagnetic field, produces a nonlinear
transport law for the pitch $\Phi$.  We refer to this branch
interchangeably as the \emph{fixed-flux} or \emph{time-independent
longitudinal-profile} branch.

The main purpose of this section is to follow the consequences of this
transport law directly.  We first show that smooth compressive pitch profiles
can produce a finite-time loss of invertibility of the radial characteristic
map.  We then identify the corresponding geometric signature in the rotation
of the normalized helical field plane and determine the Amp\`ere current
required to maintain the branch.  Finally, we discuss the entropy weak
continuation of the reduced conservation law beyond first breaking.
Throughout this section we impose \(U(1)_\phi\) symmetry.

\subsection{Helical flux-rope branch and nonlinear pitch transport}
\label{subsec:seed-deformations}
\label{subsec:seed-completion}
A helical flux-carrying simple sector with fixed radial leg can be written as
\begin{equation}
        G
        =
        dr\wedge e,
        \qquad
        e
        =
        B(r)\,d\phi+A(t,r,\chi)\,d\chi .
        \label{eq:G2}
\end{equation}
The purely longitudinal configuration
\(
G_0=B(r)\,dr\wedge d\phi
\)
is recovered for \(A=0\), whereas \(A\neq0\) introduces an azimuthal
magnetic component and hence a helical field direction.

The two-plane supporting \(G\) is generated by \(dr\) and \(e\).
Within the \((\phi,\chi)\) subspace, define the one-form
\begin{equation}
        e_\perp
        :=
        Ar^2\,d\phi-B\,d\chi .
        \label{eq:eperp}
\end{equation}
With the cylindrical metric, $g^{-1}(e,e_\perp)=0 .$
Since \(dt\) is orthogonal to both \(dr\) and \(e\), the two-plane generated
by \(dt\) and \(e_\perp\) is metric-orthogonal to the plane supporting
\(G\).
We therefore take the complementary simple sector to be
\begin{equation}
        H
        =
        \mu(t,r,\chi)\,
        dt\wedge e_\perp
        =
        \mu(t,r,\chi)\,
        dt\wedge
        \left(
        Ar^2\,d\phi-B\,d\chi
        \right).
        \label{eq:principal-partner}
\end{equation}
The scalar \(\mu\) is then determined, up to the residual time-dependent
freedom described below, by imposing the Bianchi identity~\eqref{closure} on the total field $F=G+H$.

Imposing the usual Bianchi identity \(dF=0\) gives
\begin{equation}
        \partial_r(\mu Ar^2)=0,
        \qquad
        \partial_r(\mu B)=-\partial_t A,
        \qquad
        \partial_\chi(\mu Ar^2)=0 .
        \label{eq:completion-conditions}
\end{equation}
The first and third equations imply $   \mu Ar^2=U(t),$ where \(U(t)\) is not fixed by the geometric construction itself.

It is convenient to introduce the pitch $\Phi$ and the accumulated drive
\begin{equation}
        \Phi(t,r,\chi)
        :=
        \frac{A(t,r,\chi)r^2}{B(r)},
        \qquad
        \tau(t)
        :=
        \int_{t_0}^{t}U(t')\,dt' .
        \label{eq:pitch-def}
\end{equation}
Then
\(
A=B\Phi/r^2
\)
and
\(
\mu B=U/\Phi
\),
so that the remaining equation in
Eq.~\eqref{eq:completion-conditions} reduces to
\begin{equation}
        \boxed{
        \partial_\tau\Phi
        +
        \frac{r^2}{B}
        \partial_r
        \left(
        \frac{1}{\Phi}
        \right)
        =
        0 .
        }
        \label{eom1}
\end{equation}
Equivalently, $  \partial_\tau\Phi - (r^2/(B\Phi^2))\, \partial_r\Phi =0 .$

Equation~\eqref{eom1} is the central evolution equation of the time-independent longitudinal-profile branch.  
Its characteristic velocity depends on the pitch itself.  Thus an
initially smooth configuration can develop characteristic focusing even
though the pitch amplitude remains finite.

The complementary sector takes the particularly simple form
\begin{equation}
        H
        =
        U(t)\,dt\wedge
        \left(
        d\phi-\Phi^{-1}d\chi
        \right).
        \label{H:simple}
\end{equation}
Its spatial direction has the reciprocal, oppositely signed pitch required
by metric orthogonality.  Although \(G\) and \(H\) are not closed separately,
their exterior derivatives cancel, 
$  dG=\frac{U\,\partial_r\Phi}{\Phi^2}\,dt\wedge dr\wedge d\chi =-dH,$
and hence \(dF=0\).

In the orthonormal cylindrical frame $\{e^a\} = \{dt, dr, r\,d\phi,d\chi \}$,  $(a = \hat 0,\hat r,\hat \phi, \hat \chi )$,  the electric and magnetic fields measured by the static observers are 
$\mathbf B=-A\,\hat{\boldsymbol\phi}
        +(B/r)\,\hat{\boldsymbol\chi} =-(Ar/U) \mathbf E$.
Therefore
\begin{equation}
        \boxed{
        \mathbf E\times\mathbf B=0,
        \qquad
        \mathbf E\cdot\mathbf B
        =
        -\frac{U}{Ar}
        \left(
        A^2+\frac{B^2}{r^2}
        \right)
        \neq0 .
        }
        \label{eq:nonideal-EB}
\end{equation}
The relative electric-field strength is
\begin{equation}
        \epsilon_\parallel
        :=
        \frac{|\mathbf E|}{|\mathbf B|}
        =
        \frac{|U|}{|A|r},
        \qquad
        F_{ab}F^{ab}
        =
        2|\mathbf B|^2
        \left(
        1-\epsilon_\parallel^2
        \right).
        \label{eq:nonideal-strength}
\end{equation}
The region \(\epsilon_\parallel<1\) is magnetically dominated, and
\(\epsilon_\parallel\ll1\) corresponds to a parametrically small parallel
electric field.  Nevertheless, the exact ideal-MHD condition is not restored
for any nonzero \(\epsilon_\parallel\), because the pseudoscalar
invariant \(\mathbf E\!\cdot\!\mathbf B\) in Eq.~\eqref{eq:nonideal-EB}
remains nonzero.
The branch is likewise distinct from degenerate force-free
electrodynamics, for which
\(\mathbf E\cdot\mathbf B=0\)
is part of the field constraint~\cite{GrallaJacobson2014}.

\subsection{Finite-time characteristic focusing}
\label{subsec:flux-rope-collapse-time}

To expose the characteristic structure of Eq.~\eqref{eom1}, introduce
a radial coordinate \(\sigma\) by
\begin{equation}
        d\sigma
        =
        \frac{B(r)}{r^2}\,dr .
        \label{eq:sigma-def}
\end{equation}
The pitch equation becomes the scalar conservation law
\begin{equation}
        \boxed{
        \partial_\tau\Phi
        +
        \partial_\sigma f(\Phi)
        =
        0,
        \qquad
        f(\Phi)=\frac{1}{\Phi}.
        }
        \label{eq:fixed-flux-conservation-law}
\end{equation}
The coordinate \(\chi\) enters only parametrically.  On the positive-pitch
branch, $ f''(\Phi)={2/\Phi^3}>0 ,$
so the reduced equation is a strictly convex scalar conservation law
\cite{Lax1964,Whitham1974,Dafermos2016}.

The conserved variable also has a direct magnetic interpretation.  For a
fixed interval \([\sigma_1,\sigma_2]\),
\begin{equation}
        \frac{d}{d\tau}
        \int_{\sigma_1}^{\sigma_2}\Phi\,d\sigma
        =
        -
        \left[
        \frac{1}{\Phi}
        \right]_{\sigma_1}^{\sigma_2},
        \qquad
        \Phi\,d\sigma=A\,dr .
        \label{eq:poloidal-flux-conservation}
\end{equation}
Thus, when the boundary fluxes match,
\(
\int\Phi\,d\sigma=\int A\,dr
\)
is conserved.  Up to orientation, this quantity is the poloidal magnetic
flux per unit longitudinal length through a radial--longitudinal strip.

A homogeneous state is linearly neutral.  Writing $    \Phi=\overline{\Phi}+\delta\Phi$ with $  |\delta\Phi|\ll|\overline{\Phi}|$, one obtains 
$\partial_\tau\delta\Phi-\overline{\Phi}^{-2}  \partial_\sigma\delta\Phi=0 .$
Infinitesimal perturbations are therefore advected without exponential
growth.  The finite-time breakdown found below is a nonlinear
characteristic-focusing phenomenon rather than a normal-mode instability of
the kind usually invoked for helical
ropes~\cite{TorokKliem2005,KliemTorok2006}.
The nonlinear characteristic equations are
\begin{equation}
        \frac{d\sigma}{d\tau}
        =
        -\frac{1}{\Phi^2},
        \qquad
        \frac{d\Phi}{d\tau}
        =
        0 .
        \label{characteristiceq}
\end{equation}
If \(\sigma_0\) labels the initial position of a characteristic,
\begin{equation}
        \Phi(\tau,\sigma,\chi)
        =
        \Phi_0(\sigma_0,\chi),
        \qquad
        \sigma
        =
        \sigma_0
        -
        \frac{\tau}
        {\Phi_0(\sigma_0,\chi)^2}.
        \label{eq:collapse-characteristic-map}
\end{equation}
The pitch remains finite along every characteristic, but different pitch
values propagate with different radial velocities.

The Jacobian of this map is
\begin{equation}
        \mathcal J_\sigma
        :=
        \frac{\partial\sigma}{\partial\sigma_0}
        =
        1+
        \frac{
        2\tau\,
        \partial_{\sigma_0}\Phi_0
        }{
        \Phi_0^3
        } .
        \label{eq:collapse-jacobian}
\end{equation}
As long as \(\mathcal J_\sigma\neq0\),
\begin{equation}
        \partial_\sigma\Phi
        =
        \frac{
        \partial_{\sigma_0}\Phi_0
        }{
        \mathcal J_\sigma
        } .
        \label{eq:pitch-gradient-blowup}
\end{equation}

The caustic is defined kinematically by the loss of local invertibility of
the characteristic map  $\mathcal J_\sigma=0 .$
For a generic crossing with
\(
\partial_{\sigma_0}\Phi_0\neq0
\),
the pitch gradient diverges as a consequence of this loss of invertibility.
The gradient divergence is therefore a manifestation of the caustic, not
its definition.

A local crossing occurs at
\begin{equation}
        \tau_c(\sigma_0,\chi)
        =
        -
        \frac{
        \Phi_0(\sigma_0,\chi)^3
        }{
        2\,\partial_{\sigma_0}\Phi_0(\sigma_0,\chi)
        } .
        \label{eq:local-collapse-time}
\end{equation}
A finite future crossing exists whenever
$ \partial_{\sigma_0}\Phi_0/\Phi_0^3<0 .$
For a positive-pitch profile this is simply the compressive condition
$
        \partial_{\sigma_0}\Phi_0<0 .
$
The first caustic occurs at
\begin{equation}
        \tau_\ast
        =
        \inf_{\sigma_0,\chi}
        \left\{
        -
        \frac{
        \Phi_0^3
        }{
        2\,\partial_{\sigma_0}\Phi_0
        }
        :
        \frac{
        \partial_{\sigma_0}\Phi_0
        }{
        \Phi_0^3
        }
        <0
        \right\}.
        \label{eq:first-collapse-time}
\end{equation}

\paragraph*{Domain, boundary data, and competition with the axis.}---
The characteristic statement is local unless a radial domain is specified.
Let
\(
I_r=(r_{\rm in},r_{\rm out})
\)
be an interval on which
\(
B>0
\)
and
\(
0<\Phi<\infty
\),
mapped monotonically to \(I_\sigma\) by
Eq.~\eqref{eq:sigma-def}.  Since
\(
d\sigma/d\tau<0
\),
the outer endpoint is an inflow boundary for a finite cylinder, while the
inner endpoint is outflow.  The initial-value solution applies until a
characteristic either intersects a neighbor or leaves the interval.

For the power-law profile
\(
B=b r^m
\)
with \(b>0\),
\begin{equation}
        \sigma(r)
        =
        \begin{cases}
        \dfrac{b}{m-1}r^{m-1},
        &m\neq1,\\[5pt]
        b\ln r,
        &m=1.
        \end{cases}
        \label{eq:sigma-power-law}
\end{equation}
For \(m>1\) the additive constant can be chosen so that the axis is at
\(\sigma=0\); for \(m\leq1\) it lies at infinite \(\sigma\)-distance.
A positive-pitch characteristic beginning at \(\sigma_0>0\) reaches the
axis at $\tau_{\rm ax}(\sigma_0)=\sigma_0\Phi_0(\sigma_0)^2 .$
Thus, when the axis is at finite characteristic distance, an interior caustic
is encountered first provided
\begin{equation}
        \tau_c(\sigma_0)
        <
        \tau_{\rm ax}(\sigma_0)
        \quad\Longleftrightarrow\quad
        \frac{\Phi_0}
        {2|\partial_{\sigma_0}\Phi_0|}
        <
        \sigma_0 ,
        \qquad
        \partial_{\sigma_0}\Phi_0<0 .
        \label{eq:caustic-before-axis}
\end{equation}

The accumulated variable separates the geometric threshold from its
realization in physical time:
\begin{equation}
        \int_{t_0}^{t_\ast}
        U(t')\,dt'
        =
        \tau_\ast .
        \label{eq:physical-collapse-time}
\end{equation}
Consequently, a finite pulse remains pre-caustic if its total accumulated
drive is smaller than \(\tau_\ast\), whereas an arbitrarily weak but
sufficiently persistent drive can reach the same geometric threshold.
A constitutive upper bound on the admissible electric field can change the
physical time required to accumulate \(\tau_\ast\), but not the geometric
threshold itself; see Appendix~\ref{app:physical-scales}.

\subsection{Geometric signature: rotation of the normalized helical plane}
\label{subsec:geometric-signature}

The characteristic caustic is a statement about the radial map.  We now ask
how this loss of invertibility appears intrinsically in the local geometry of
the helical field.

Remove the scalar magnitude of the one-form \(e\) by defining
\begin{equation}
        \mathcal N^2
        :=
        g^{-1}(e,e)
        =
        \frac{B^2}{r^2}+A^2,
        \qquad
        \widehat e
        :=
        \frac{e}{\mathcal N},
        \qquad
        \widehat G
        :=
        dr\wedge\widehat e .
        \label{eq:normalized-support}
\end{equation}
The orientation of the normalized helical direction is conveniently
described by the angle $  \tan\vartheta :=Ar/B= \Phi/r.$
The local physical-time rotation rate of the normalized field plane is
therefore \(\partial_t\vartheta\).

The same quantity has an invariant differential-form expression.  Within
the fixed-radial-leg family, Frobenius integrability of the kernel of
\(\widehat G\)~\cite{BryantEDS1991} is measured by
\(
\widehat e\wedge d\widehat G
\).
We define
\begin{equation}
        \mathcal A_{\rm F}[G]
        :=
        -\star
        \left(
        \widehat e\wedge d\widehat G
        \right)
        =
        -
        \frac{
        \star(e\wedge dG)
        }{
        \mathcal N^2
        } .
        \label{eq:anholonomy-def}
\end{equation}
The sign is fixed so that positive
\(\mathcal A_{\rm F}\) corresponds to increasing \(\vartheta\) with the
orientation adopted above.

For the axisymmetric sector, allowing temporarily for time dependence of
both \(A\) and \(B\),
\begin{equation}
        \mathcal A_{\rm F}[G]
        =
        \frac{
        B\,\partial_tA
        -
        A\,\partial_tB
        }{
        r\left(B^2/r^2+A^2\right)
        }
        =
        \partial_t\vartheta
       = \frac{  \partial_t\Phi}{ r(1+\Phi^2/r^2)} ,
        \label{eq:anholonomy-value}
\end{equation}
where we use \(\tan\vartheta=\Phi/r\) in the third equality.
Thus the normalized Frobenius scalar has a direct physical interpretation:
it is simply the local rotation rate of the normalized helical field plane.

For the present branch with a time-independent longitudinal profile
\(B=B(r)\), using
\(
\partial_t\Phi=U\,\partial_\tau\Phi
\)
together with Eq.~\eqref{eom1} gives
\begin{equation}
        \boxed{
        \mathcal A_{\rm F}[G]
        =
        \frac{Ur}{B}
        \frac{
        \partial_r\Phi
        }{
        \Phi^2
        \left(
        1+\Phi^2/r^2
        \right)
        } .
        }
        \label{eq:anholonomy-pitch}
\end{equation}
Hence, provided
\(
B,U,r,\Phi
\)
remain finite and nonzero at the first crossing,
\begin{equation}
        \boxed{
        \mathcal J_\sigma
        \longrightarrow0
        \quad\Longrightarrow\quad
        |\partial_r\Phi|
        \longrightarrow\infty
        \quad\Longrightarrow\quad
        |\mathcal A_{\rm F}[G]|
        \longrightarrow\infty .
        }
        \label{eq:crossing-anholonomy-chain}
\end{equation}
The hierarchy is important.  The caustic itself is defined by
\(
\mathcal J_\sigma=0
\),
that is, by loss of invertibility of the characteristic map.
The divergence of
\(
\mathcal A_{\rm F}
\)
is the intrinsic geometric signature of that collapse within the regular
helical branch.  The pitch-gradient divergence is the particular way in
which this signature is realized in the fixed-flux solution.

\begin{remark}[Local blow-up rate and finite angle] \label{rem:blow-up-rate}
For an ordinary nondegenerate first breaking point of a smooth scalar
conservation law \cite{Whitham1974,Dafermos2016}, the Eulerian profile has the cubic-root form
$\Phi-\Phi_c=O(|\sigma-\sigma_c|^{1/3})$ at $\tau=\tau_c$, so
$\partial_\sigma\Phi=O(|\sigma-\sigma_c|^{-2/3})$.  At the fixed breaking
location the corresponding time singularity is
$|\mathcal A_{\rm F}|=O((\tau_c-\tau)^{-2/3})$.  The exponent is integrable:
the plane angle $\vartheta=\arctan(\Phi/r)$ remains finite even though its
Eulerian rotation rate diverges.  The catastrophe is therefore a singular
rotation rate, not infinite winding of the helical field plane.
\end{remark}

\subsection{Electromagnetic signature: a current-sheet precursor}
\label{subsec:current-sheet-precursor}

The characteristic caustic and the divergent field-plane rotation rate are differential singularities.  
They do not by themselves imply that the electromagnetic field amplitude becomes infinite.  
To determine the physical electromagnetic response, we evaluate the invariants and the Amp\`ere current
of the complete field \(F=G+H\).

For Eqs.~\eqref{eq:G2} and \eqref{H:simple}, with
\(
A=B\Phi/r^2
\),
the algebraic Maxwell invariants are
\begin{align}
        F_{ab}F^{ab}
        &=
        2
        \left(
        1+\frac{\Phi^2}{r^2}
        \right)
        \left(
        \frac{B^2}{r^2}
        -
        \frac{U^2}{\Phi^2}
        \right),
        \qquad
        F_{ab}{}^\star F^{ab}
        =
        \frac{4UB}{r\Phi}
        \left(
        1+\frac{\Phi^2}{r^2}
        \right).
        \label{eq:fixed-flux-invariant-2}
\end{align}
Neither invariant contains a derivative of \(\Phi\).  Hence, if
\(
B,U,\Phi,\Phi^{-1}
\)
remain finite at first crossing, the divergence of
\(
\partial_r\Phi
\)
does not force the field or its algebraic invariants to diverge.
Moreover, for \(UB\Phi\neq0\),
\(
F_{ab}{}^\star F^{ab}\neq0
\),
so the field remains nondegenerate even if
\(
F_{ab}F^{ab}
\)
passes through zero.

The physical current is $J^\mu=\nabla_\nu F^{\nu\mu}.$
Using Eq.~\eqref{eom1}, its longitudinal component can be written as
\begin{equation}
        J^\chi
        =
        \frac{\dot U}{\Phi}
        +
        \frac{B'\Phi}{r^2}
        -
        \frac{B\Phi}{r^3}
        +
        \left(
        \frac{B}{r^2}
        -
        \frac{U^2r^2}{B\Phi^4}
        \right)
        \partial_r\Phi .
        \label{eq:fixed-flux-current-chi}
\end{equation}

Introduce the physical radial velocity of a pitch characteristic,
\begin{equation}
        v_{\rm c}
        :=
        \left.
        \frac{dr}{dt}
        \right|_{\rm char}
        =
        -
        \frac{Ur^2}{B\Phi^2}.
        \label{eq:physical-characteristic-velocity}
\end{equation}
The complementary sector can equivalently be written as
\(
H=h\star G
\)
with
\(
h=-U/(Ar)
\),
which is the notation used from Sec.~\ref{sec:radial-extension} onward; note
that $|h|=\epsilon_\parallel$ by Eq.~\eqref{eq:nonideal-strength}.
Since
\(
\tan\vartheta=Ar/B
\),
the characteristic speed may also be written as
\begin{equation}
        v_{\rm c}
        =
        h\cot\vartheta,
        \qquad
        |v_{\rm c}|
        =
        \epsilon_\parallel
        |\cot\vartheta|.
        \label{eq:vc-h-epsilon}
\end{equation}

The part of the current that carries the singular pitch gradient is
\begin{equation}
        \boxed{
        J^\chi_{\rm sing}
        =
        \frac{B}{r^2}
        \left(
        1-v_{\rm c}^2
        \right)
        \partial_r\Phi .
        }
        \label{eq:current-singular-part}
\end{equation}
Thus a generic characteristic crossing produces a divergent physical
current provided the prefactor remains nonzero.  More precisely, if
$\liminf_{t\to t_c}   |1-v_{\rm c}^2|>0 ,$ then the inverse-Jacobian growth of the pitch gradient is inherited directly by \(J^\chi\).

Indeed, Eqs.~\eqref{eq:sigma-def} and \eqref{eq:pitch-gradient-blowup} give
$ \partial_r\Phi
        =(B/r^2) (\partial_{\sigma_0}\Phi_0 /\mathcal J_\sigma).$
Hence a generic first crossing obeys
\begin{equation}
        \boxed{
        J^\chi
        \sim
        \mathcal J_\sigma^{-1}
        \qquad
        \text{as}
        \qquad
        \mathcal J_\sigma\to0 .
        }
        \label{eq:current-sheet-scaling}
\end{equation}
To make the cancellation explicit, decompose the Amp\`ere current according to
\begin{equation}
        J^\mu
        =
        J_G^\mu+J_H^\mu,
        \qquad
        J_G^\mu:=\nabla_\nu G^{\nu\mu},
        \qquad
        J_H^\mu:=\nabla_\nu H^{\nu\mu}.
\end{equation}
Near the caustic, the singular longitudinal contributions satisfy
$ J_{G,{\rm sing}}^\chi=\frac{B}{r^2}\partial_r\Phi$ and 
$ J_{H,{\rm sing}}^\chi=-v_{\rm c}^2 J_{G,{\rm sing}}^\chi .$
Hence $ J_{\rm sing}^\chi
        = \left(1-v_{\rm c}^2\right)
        J_{G,{\rm sing}}^\chi ,$
where a direct current-locking interpretation for the $1-v_{\rm c}^2$ factor is given in Ref.~\cite{Kim2}.   
At the special value \(|v_{\rm c}|=1\), the two singular contributions are
equal in magnitude and opposite in sign, so their leading divergence cancels
in the physical current \(J^\chi\).  The separate contributions
\(J_G^\chi\) and \(J_H^\chi\), however, remain singular when
\(\partial_r\Phi\) diverges.

The electromagnetic work density measured by the static observers has the
same singular factor.  Since $J^{\hat\phi}=rJ^\phi$,
$E_{\hat\phi}=U/r$, and $E_{\hat\chi}=-U/\Phi$, its singular part is
\begin{equation}
        (\mathbf J\cdot\mathbf E)_{\rm sing}
        =
        -\frac{UB}{r^2\Phi}
        \left(
        1-v_{\rm c}^2
        \right)
        \partial_r\Phi .
        \label{eq:JE-singular-part}
\end{equation}
This is the Maxwell work density required by the prescribed branch.
Calling it ``Ohmic heating'' would require an additional constitutive
relation; no such resistive interpretation is assumed here.

For a strictly \(\chi\)-independent configuration,
\(
J^t=J^r=0
\)
and \(J^\phi\) remains regular, so the singular current is carried by
\(J^\chi\).  With transverse pitch dependence, the same inverse-Jacobian
amplification can appear in additional components.

The result is therefore conditional but sharp: if the helical branch remains
valid up to first characteristic crossing and the current prefactor does not
vanish, the electromagnetic field can remain finite while the current
required to support that branch diverges.  We interpret this formal endpoint
as a \emph{current-sheet precursor}.

\begin{proposition}[Fixed longitudinal-profile caustic]
\label{prop:fixed-flux-caustic}
Let \(B>0\) on a radial interval and let \(\Phi_0>0\) be \(C^2\).
Suppose there exists a compressive label satisfying
\(
\partial_{\sigma_0}\Phi_0<0
\)
whose characteristic remains in the nondegenerate helical branch until its
local crossing time~\eqref{eq:local-collapse-time}.  If the axis is at finite
\(\sigma\)-distance, assume also Eq.~\eqref{eq:caustic-before-axis}.
Then the characteristic map loses local invertibility at finite accumulated
time,
\(
\mathcal J_\sigma\to0
\),
while the pitch remains finite.
If \(B,U,r,\Phi,\Phi^{-1}\) remain finite and nonzero at the crossing, the
normalized field-plane rotation rate
\(
|\mathcal A_{\rm F}|
\)
diverges.  If, in addition, the prefactor in
Eq.~\eqref{eq:current-singular-part} remains nonzero, the Amp\`ere current
required to support this electromagnetic branch diverges as well.
\end{proposition}
No constitutive matter model is assumed in the proposition.  The current
\(J^\mu=\nabla_\nu F^{\nu\mu}\) is the source required by Maxwell's equations
for the prescribed electromagnetic branch.  Whether a self-consistent plasma
can supply this current and remain on the branch up to the caustic is a
separate matter--field consistency question, considered in
Sec.~\ref{sec:matter-closure}.

\subsection{Entropy weak continuation beyond first breaking}
\label{subsec:entropy-continuation}

The loss of the classical characteristic solution at
\(
\mathcal J_\sigma=0
\)
does not end the reduced scalar conservation law
\eqref{eq:fixed-flux-conservation-law}.  Within that reduced equation, the
solution can be continued in the standard weak sense.  Since
\(
f(\Phi)=1/\Phi
\)
is strictly convex for \(\Phi>0\), the entropy admissibility condition
selects the corresponding compressive shock
\cite{Lax1964,Whitham1974,Dafermos2016}.
Here ``entropy'' refers to entropy admissibility in the mathematical theory
of scalar conservation laws.  It does not by itself constitute a
thermodynamic entropy-production law for the plasma.

Let the shock separate positive-pitch states
\(
\Phi_L>\Phi_R>0
\),
and define
\(
[X]:=X_R-X_L
\).
The Rankine--Hugoniot relation gives
\begin{equation}
        \boxed{
        \frac{d\sigma_s}{d\tau}
        =
        \frac{[1/\Phi]}{[\Phi]}
        =
        -\frac{1}{\Phi_L\Phi_R}.
        }
        \label{eq:RH-pitch-shock}
\end{equation}
For the convex flux \(f\), this compressive jump satisfies the scalar Lax
entropy condition.

In physical radius, the corresponding propagation speed is
\begin{equation}
        V_s
        :=
        \frac{dr_s}{dt}
        =
        -
        \frac{
        U(t)r_s^2
        }{
        B(r_s)\Phi_L\Phi_R
        } .
        \label{eq:physical-shock-speed}
\end{equation}

The weak continuation also has a distributional electromagnetic signature.
Across the moving jump, the singular part of \(J^\chi\) is
$J^\chi_{\rm sheet}= \left( [A]- V_sU[1/\Phi] \right)\delta(r-r_s).$
Using
\(
A=B\Phi/r_s^2
\)
and Eq.~\eqref{eq:RH-pitch-shock},
\begin{equation}
        \boxed{
        J^\chi_{\rm sheet}
        =
        \frac{B(r_s)}{r_s^2}
        \left(
        1-V_s^2
        \right)
        [\Phi]\,
        \delta(r-r_s),
        }
        \label{eq:surface-current-shock}
\end{equation}
up to the overall sign associated with the jump convention.
The factor \(1-V_s^2\) is the distributional counterpart of the same
cross-sector cancellation appearing in the smooth precursor.
Thus, within the reduced conservation law, the classical caustic has a
natural entropy-shock continuation carrying a surface Amp\`ere current.

This statement should be distinguished from the dynamical selection of a
finite-width current layer in a self-consistent plasma.  The reduced
conservation law determines its own entropy weak solution once first breaking
has occurred, but it does not determine whether the physical plasma reaches
that regime or whether resistive, Hall, kinetic, or dispersive physics selects
the same microscopic current-layer structure.  Those questions require an
additional material model.  The present result instead identifies the
geometric route from characteristic focusing to current concentration~\cite{Grad1967,Parker1972,NgBhattacharjee1998,PontinHornig2020}.

\paragraph*{Example.}---
As a concrete illustration, take $B=r^2$, so that $ \sigma=r$, on \(0<r<4\), with initial pitch
\begin{equation}
        \Phi_0(\sigma)
        =
        1.4
        -
        0.4
        \tanh
        \left(
        \frac{\sigma-2}{0.35}
        \right).
        \label{eq:numerical-profile}
\end{equation}
Direct minimization of
Eq.~\eqref{eq:local-collapse-time} gives $ \sigma_{0\ast}\simeq2.150$,
$ \Phi_{0\ast}\simeq1.238$, and $ \tau_\ast\simeq0.993 .$
The first crossing occurs at $  \sigma_\ast\simeq1.502,$ whereas the critical characteristic reaches the axis only at $ \tau_{\rm ax}\simeq3.296 .$
The first caustic is therefore an interior event rather than an artifact of
the axis.

\begin{figure*}[t]
\centering
\includegraphics[width=0.47\textwidth]{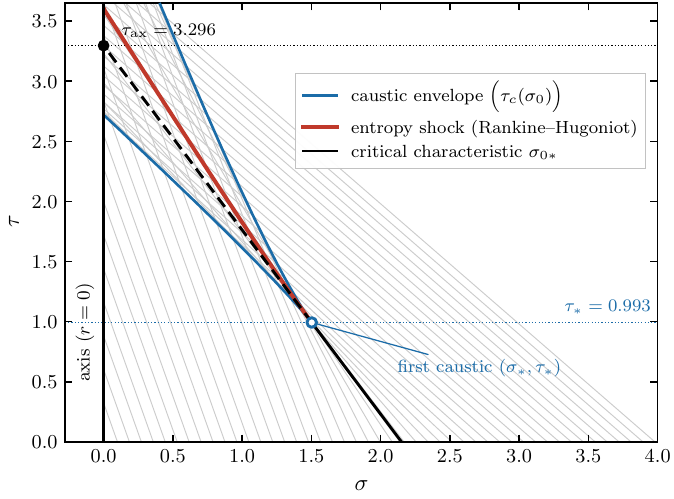}\hfill
\includegraphics[width=0.47\textwidth]{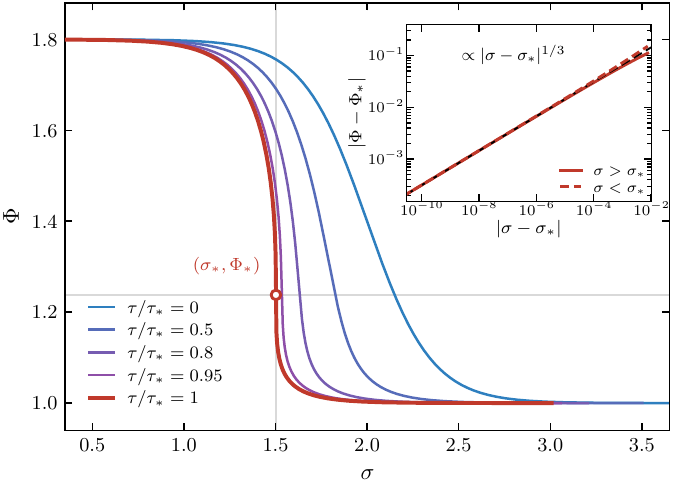}
\caption{
Time-independent longitudinal-profile focusing~\eqref{eq:numerical-profile} with \(B=r^2\) and \(\sigma=r\).
\textbf{(a)}
The characteristic fan first loses invertibility at
\((\sigma_\ast,\tau_\ast)\simeq(1.502,0.993)\), well before the critical
characteristic reaches the axis at
\(\tau_{\rm ax}\simeq3.296\).
The red curve shows the entropy-shock continuation selected by the
Rankine--Hugoniot and Lax conditions for the reduced scalar conservation law.
\textbf{(b)}
The pitch remains within its initial finite range while its gradient
steepens.
At first breaking the profile has the cubic-root behavior
\(
|\Phi-\Phi_\ast|
\propto
|\sigma-\sigma_\ast|^{1/3}
\),
as discussed in Remark~\ref{rem:blow-up-rate}.
}
\label{fig:fixed-flux-example}
\end{figure*}

Figure~\ref{fig:fixed-flux-example} shows the characteristic fan, the first
caustic, the entropy-shock continuation of the reduced conservation law, and
the associated pitch steepening.

\section{Radial magnetic through-flux and robustness of the focusing}
\label{sec:radial-extension}

The caustic obtained in Sec.~\ref{sec:helical-fluxtube} was derived for a
helical sector whose one-form paired with \(dr\) was purely spatial.  We now
ask whether the focusing can be removed by allowing radial magnetic
through-flux while preserving the radial organization of the flux rope.

Keeping \(dr\) as one leg, the most general simple sector within this
fixed-radial-leg class is
\begin{equation}
        G
        =
        dr\wedge
        \left(
        P\,dt+B\,d\phi+A\,d\chi
        \right).
        \label{eq:general-radial-leg-sector}
\end{equation}
The temporal component \(P\,dt\) tilts the support plane of \(G\) into the
time direction without changing its radial leg.  As shown below, metric
orthogonality then introduces a \(d\phi\wedge d\chi\) component in the
complementary sector, corresponding to magnetic flux through a surface of
constant \(r\).

For $ \Delta
        :=
        \frac12 G_{ab}G^{ab}
        =
        A^2+{B^2/r^2}-P^2
        \neq0,$
the sector is non-null.  Its metric-orthogonal complementary two-plane is
the support plane of \(\star G\).  With the orientation adopted in
Sec.~\ref{sec:intro}, $  \star G
        =
        Pr\,d\phi\wedge d\chi
        +
        \frac{B}{r}\,dt\wedge d\chi
        -
        Ar\,dt\wedge d\phi .$
We therefore write the complementary simple sector as
\begin{equation}
        H
        =
        h\,\star G
        =
        h\left[
        Pr\,d\phi\wedge d\chi
        +
        \frac{B}{r}\,dt\wedge d\chi
        -
        Ar\,dt\wedge d\phi
        \right].
        \label{eq:radial-extension-partner}
\end{equation}
The limit \(P\to0\) smoothly recovers the complementary sector used in
Sec.~\ref{sec:helical-fluxtube}.

The new spatial term motivates the definitions
\begin{equation}
        Q:=hPr,
        \qquad
        U:=-hAr .
        \label{eq:radial-QU}
\end{equation}
Here \(Q\) is the local density of radial magnetic flux through the
fixed-\(r\) torus \(T_r^2\) parametrized by \((\phi,\chi)\).  Explicitly,
\begin{equation}
        \Psi_r(t,r)
        :=
        \int_{T_r^2}H
        =
        2\pi
        \int_0^{L_\chi} Q(t,r,\chi)\,d\chi .
        \label{eq:radial-through-flux}
\end{equation}
The quantity \(U\) reduces to the pitch-transport amplitude of the
zero-through-flux branch.

\subsection{Closure equations and radial-flux transport}
\label{subsec:radial-flux-closure}

We retain the \(U(1)_\phi\) symmetry used in the previous section.
The Bianchi identity for \(F=G+H\) gives
\begin{equation}
\label{eq:radial-extension-closure}
        \partial_tB+\partial_r(hAr)=0, \quad
        \partial_tA-\partial_\chi P
        -\partial_r\!\left(\frac{hB}{r}\right)=0, \quad
        \partial_tQ-\partial_\chi(hAr)=0, \quad
        \partial_\chi B+\partial_rQ=0.
\end{equation}
Using Eq.~\eqref{eq:radial-QU}, three of these equations become
\begin{equation}
        \partial_rU=\partial_tB,
        \qquad
        \partial_tQ=-\partial_\chi U,
        \qquad
        \partial_rQ=-\partial_\chi B .
        \label{eq:radial-extension-QU-system}
\end{equation}

These relations make the role of the new degree of freedom transparent.
For periodic \(\chi\),
\begin{equation}
        \partial_t
        \int_0^{L_\chi}Q\,d\chi
        =
        0,
        \qquad
        \partial_r
        \int_0^{L_\chi}Q\,d\chi
        =
        0 .
        \label{eq:radial-extension-average}
\end{equation}
Thus the net radial magnetic flux through \(T_r^2\) is conserved, whereas
its local density \(Q(t,r,\chi)\) may be redistributed along the
longitudinal direction.

\subsection{Time-independent longitudinal profile}
\label{subsec:radial-flux-fixed-profile}

We now make the comparison with Sec.~\ref{sec:helical-fluxtube} by keeping
the longitudinal magnetic profile time independent, $B=B(r)$ with $ \partial_tB=0$.
Equation~\eqref{eq:radial-extension-QU-system} then gives $\partial_rU=0$,
so \(U=U(t,\chi)\).  Introducing again the pitch~\eqref{eq:pitch-def}, the second equation in Eq.~\eqref{eq:radial-extension-closure} becomes
\begin{equation}
        \boxed{
        \partial_t\Phi
        -
        \frac{Ur^2}{B\Phi^2}\,
        \partial_r\Phi
        =
        \frac{r^2}{B}\,
        \partial_\chi P .
        }
        \label{eq:radial-extension-pitch}
\end{equation}
The radial through-flux therefore modifies the pitch transport through the
longitudinal variation of \(P\).

The simplest and most direct test is the \(\chi\)-independent sector.  In
that case $ \partial_\chi A
        =
        \partial_\chi B
        =
        \partial_\chi P
        =
        \partial_\chi h
        =
        0 ,$
and Eq.~\eqref{eq:radial-extension-QU-system} gives
\begin{equation}
        Q=q_m=\mathrm{const.},
        \qquad
        \frac{Q}{U}
        =
        -\frac{P}{A},
        \qquad
        P
        =
        -\frac{q_m}{U}\,A .
        \label{eq:radial-flux-constant}
\end{equation}
in both space and time.  The total radial magnetic flux is consequently
$  \Psi_r=2\pi L_\chi q_m .$

Because \(\partial_\chi P=0\), the source term on the right-hand side of
Eq.~\eqref{eq:radial-extension-pitch} vanishes.  With $ \tau(t)=\int_{t_0}^{t}U(t')\,dt',$ the pitch equation exactly reproduces the transport equation obtained in Sec.~\ref{sec:helical-fluxtube}.

The conclusion is therefore stronger than merely saying that the radial
flux is a small perturbation.  A finite, conserved radial through-flux can
be present while the radial characteristic equation, its collapse threshold,
and the first-caustic mechanism remain unchanged.  In particular, for the
same compressive initial pitch profile, the characteristic map reaches the
same loss of invertibility as in the zero-through-flux branch.

\subsection{Longitudinal redistribution and limits of the extension}
\label{subsec:radial-flux-redistribution}

When \(\chi\)-dependence is allowed,
\(\partial_\chi P\) acts as a source in
Eq.~\eqref{eq:radial-extension-pitch}.  The radial-flux density can then be
redistributed into regions of concentration and depletion while its
\(\chi\)-averaged value remains fixed by
Eq.~\eqref{eq:radial-extension-average}.  Such redistribution can modify the
local pitch evolution and may corrugate or filament an otherwise symmetric
approach to the caustic.

The closure equations alone, however, contain no intrinsic diffusive,
dispersive, or gradient-stiffness term that would force this redistribution
to smooth the radial focusing.  Thus longitudinal variation of the radial
through-flux provides an additional kinematic channel, but not a generic
regularization mechanism.  Whether it actually prevents characteristic
crossing requires additional constitutive dynamics.

The nonzero-flux branch also has its own algebraic limitation.  In the
\(\chi\)-independent sector,
Eq.~\eqref{eq:radial-flux-constant} gives
\begin{equation}
        \Delta
        =
        A^2
        \left(
        1-\frac{q_m^2}{U^2}
        \right)
        +
        \frac{B^2}{r^2}.
        \label{eq:radial-extension-null-condition}
\end{equation}
The non-null condition \(\Delta\neq0\) can therefore fail before the radial
caustic is reached.  Moreover, for \(q_m\neq0\) and finite nonzero \(A\),
Eq.~\eqref{eq:radial-flux-constant} prevents a smooth passage through
\(U=0\).  These are additional limits of the radial-through-flux branch,
rather than mechanisms that remove the original focusing.

We conclude that radial magnetic through-flux does not by itself stabilize
the flux rope within the fixed-radial-leg class.  In the
\(\chi\)-independent sector the original radial focusing survives exactly on
the sub-branch for which the sector nondegeneracy
\(\Delta>0\), that is \(q_m^2<U^2(1+B^2/A^2r^2)\), holds up to the caustic;
with longitudinal dependence, the new degree of freedom redistributes flux
but supplies no intrinsic smoothing scale.

This result does not exclude regular configurations outside the present
structural class.  If the radial leg is itself allowed to tilt, schematically
\(dr\rightarrow dr+C\,dt\), a broader double-deformed construction admits an
externally supported stationary helical branch in the presence of a
nonvanishing background electric field~\cite{Kim:double}.  Such a
configuration changes the local support planes of the flux rope and is
therefore a distinct electromagnetic branch, rather than stabilization of
the present rope by radial through-flux.

\section{Time-dependent longitudinal profile and geometric regularity}
\label{sec:time-dependent-B}
The radial-flux extension shows that enlarging the plane of $G$ within the
fixed-radial-leg flux-rope class does not by itself remove the characteristic
focusing.  We therefore return to the zero-through-flux branch, $P=0$, and
allow the longitudinal magnetic profile $B$ itself to evolve in time.  This
changes both the radial characteristic speed and the pitch carried by each
characteristic, and therefore provides a genuine channel through which the
focusing may be delayed or avoided.

The two orthogonal simple sectors take the form
\begin{equation}
        G
        =
        dr\wedge
        \left(
        B\,d\phi
        +
        A\,d\chi
        \right), \qquad
        H
        =
        h
        \left(
        \frac{B}{r}\,dt\wedge d\chi
        -
        Ar\,dt\wedge d\phi
        \right).
        \label{eq:H-P-zero}
\end{equation}
The Bianchi closure~\eqref{closure} reduces to
\(
\partial_t B+\partial_r(hAr)=0
\)
and
\(
\partial_t A-\partial_r(hB/r)=0
\).
The remaining components require
\(\partial_\chi B=0\) and \(\partial_\chi(hAr)=0\).  Hence
\begin{equation}
        B=B(t,r),
        \qquad
        -hAr=U(t,r).
\end{equation}
Then $h=-U/(Ar)$, and the complementary sector takes the same form as
Eq.~\eqref{H:simple}, now with $U=U(t,r)$, while the pitch $\Phi$ is
\(
\Phi:=Ar^2/B
\)
as in Eq.~\eqref{eq:pitch-def}.  In terms of $U$, the closure equations become
\begin{equation}
        \boxed{
        \partial_t B
        =
        \partial_r U,
        \qquad
        \partial_t A
        +
        \partial_r
        \left(
        \frac{U}{\Phi}
        \right)
        =
        0.
        }
        \label{eq:Bdot-Ur}
\end{equation}
Thus a time-dependent longitudinal profile is accompanied by radial
inhomogeneity of $U$,
\(
\partial_tB\neq0\Longleftrightarrow\partial_rU\neq0
\).

The second relation in Eq.~\eqref{eq:Bdot-Ur} is the same Faraday component
that generated the time-independent longitudinal-profile transport law.
Metric orthogonality fixes
\(
E_{\hat\chi}=-U/\Phi=-UB/(Ar^2)
\),
so the Maxwell--Bianchi evolution of the azimuthal field
\(B_{\hat\phi}=-A\)
contains the reciprocal nonlinearity $1/A$.  In the present branch, the
orthogonal field construction therefore supplies the algebraic relation that
closes the pitch evolution without introducing an independent ideal-MHD Ohm
law.

\subsection{Generalized pitch equation}
\label{subsec:generalized-pitch-equation}
Equation~\eqref{eq:Bdot-Ur} determines only the radial derivative of $U$.
For a prescribed $B(t,r)$, one must also specify the value of $U$ at one
radial reference surface,
\(
U_\ast(t):=U(t,r_\ast)
\).
We define the accumulated time and scaled amplitude
\begin{equation}
        \tau(t)
        :=
        \int_0^t U_\ast(t')\,dt',
        \qquad
        \mathcal U(\tau,r)
        :=
        \frac{U(t(\tau),r)}{U_\ast(t(\tau))},
        \qquad
        \mathcal U(\tau,r_\ast)=1.
\end{equation}
The reparametrization is valid on any time interval on which
$U_\ast(t)\neq0$.  The first equation of Eq.~\eqref{eq:Bdot-Ur} becomes
\(
\partial_r\mathcal U=\partial_\tau B
\),
so that
\begin{equation}
        \mathcal U(\tau,r)
        =
        1+
        \int_{r_\ast}^{r}
        \partial_\tau B(\tau,s)\,ds,
        \qquad
        U(t,r)
        =
        U_\ast(t)\,\mathcal U(\tau(t),r).
        \label{eq:U-from-B}
\end{equation}

Using $A=B\Phi/r^2$ and
\(\partial_\tau B=\partial_r\mathcal U\)
in the second equation of Eq.~\eqref{eq:Bdot-Ur}, the pitch equation becomes
\begin{equation}
        \boxed{
        \partial_\tau \Phi
        -
        K\,
        \frac{\partial_r\Phi}{\Phi^2}
        =
        -q
        \left(
        \Phi
        +
        \frac{r^2}{\Phi}
        \right).
        }
        \label{eq:generalized-pitch-Kq}
\end{equation}
Here $ K(\tau,r)
        :=r^2\mathcal U/B $ and $ q(\tau,r)
        := {\partial_\tau B/B}
        ={\partial_r\mathcal U/ B}.$
Once the reference surface $r_\ast$ and accumulated time $\tau$ are fixed,
the reduced pitch dynamics is determined by the single scalar profile
$B(\tau,r)$.  The function $U_\ast(t)$ enters only through the reconstruction
of physical time.  We choose the orientation of accumulated time so that
$U_\ast(t)>0$.  With this convention, $q\leq0$ corresponds to relaxation of
the longitudinal profile in physical time.

The time-independent longitudinal-profile limit is recovered when
\(\partial_r\mathcal U=0\).  Equation~\eqref{eq:Bdot-Ur} then gives
\(\partial_tB=0\), and Eq.~\eqref{eq:generalized-pitch-Kq} reduces to
Eq.~\eqref{eom1}.

\subsection{Characteristic evolution and pitch amplitude}
\label{subsec:generalized-characteristics}
The evolving longitudinal profile introduces two distinct regularity
questions: whether the pitch itself approaches a limiting amplitude, and
whether neighboring radial characteristics lose local invertibility.  We
first examine the pitch evolution along the characteristics.

The characteristic equations associated with
Eq.~\eqref{eq:generalized-pitch-Kq} are
\begin{equation}
        \boxed{
        \frac{dr}{d\tau}
        =
        -\frac{K}{\Phi^2}
        =
        -\frac{r^2\mathcal U}{B\Phi^2},
        \qquad
        \frac{d\Phi}{d\tau}
        =
        -q
        \left(
        \Phi
        +
        \frac{r^2}{\Phi}
        \right).
        }
        \label{eq:characteristic-Phi-general}
\end{equation}
For \(\partial_\tau B=0\), one has $q=0$, and $\Phi$ is conserved along
each characteristic.  When \(\partial_\tau B\neq0\), the pitch itself
evolves along the characteristic flow.

The second characteristic equation becomes the linear equation
\begin{equation}
        \frac{dY}{d\tau}
        =
        -2q\left(Y+r^2\right),
        \qquad
        Y:=\Phi^2.
        \label{eq:Y-characteristic}
\end{equation}
Its solution is
\begin{equation}
        Y(\tau)
        =
        e^{-2\mathcal Q(\tau)}
        \left[
        Y_0
        -
        2
        \int_0^\tau
        e^{2\mathcal Q(s)}
        q\bigl(s,r(s)\bigr)
        r(s)^2\,ds
        \right],
        \label{eq:Y-solution}
\end{equation}
where $  \mathcal Q(\tau):=\int_0^\tau q\bigl(s,r(s)\bigr)\,ds.$
If $q\leq0$ along a characteristic, an initially positive $Y_0$ cannot
reach zero.  A sufficiently large positive accumulated value of $q$, by
contrast, can drive the pitch toward the lower boundary $\Phi=0$.  Since
Eq.~\eqref{eq:Y-characteristic} is linear in $Y$, finite initial $Y$ cannot
blow up at finite $\tau$ while $q$ and $r$ remain bounded.  Thus
$|\Phi|\to\infty$ is not a generic finite-time outcome of a regular
longitudinal-profile history.

The pitch boundary $\Phi=0$ should not be identified with a caustic.  If
$U$ remains nonzero, the complementary-sector amplitude diverges there.  If
$U$ vanishes at a compensating rate, the field amplitudes can instead remain
finite and the solution may leave the helical interior regularly.  In either
case, the caustic question is controlled independently by the characteristic
Jacobian introduced next.

\subsection{Characteristic caustic and the Jacobi map}
\label{subsec:characteristic-crossing-general}
Let
\(
r=r(\tau;r_0)
\)
and
\(
\widetilde\Phi(\tau;r_0)
:=\Phi\bigl(\tau,r(\tau;r_0)\bigr)
\)
denote the characteristic trajectory and the pitch evaluated along it, where
$r_0$ is the initial radial label,
\(
r(0;r_0)=r_0
\)
and
\(
\widetilde\Phi(0;r_0)=\Phi_0(r_0)
\).
Define the Jacobi variables
\begin{equation}
        \mathcal R(\tau;r_0)
        :=
        \frac{\partial r(\tau;r_0)}{\partial r_0},
        \qquad
        \mathcal S(\tau;r_0)
        :=
        \frac{\partial\widetilde\Phi(\tau;r_0)}{\partial r_0}.
        \label{eq:Jacobi-variables}
\end{equation}
Their initial conditions are
\(
\mathcal R(0;r_0)=1
\)
and
\(
\mathcal S(0;r_0)=\Phi_0'(r_0)
\).
For time-independent $B$, the radial and $\sigma$-space Jacobians are related
by
\begin{equation}
        \mathcal J_\sigma
        =
        \mathcal R\,
        \frac{B(r)/r^2}{B(r_0)/r_0^2},
        \label{eq:J-sigma-R-relation}
\end{equation}
so their zeros coincide away from degeneracies of $B/r^2$.  The variable
$\mathcal R$ is therefore the natural extension of the crossing criterion
when $B$ itself evolves.

As long as the characteristic map remains invertible,
\begin{equation}
        \left.
        \partial_r\Phi(\tau,r)
        \right|_{r=r(\tau;r_0)}
        =
        \frac{\mathcal S(\tau;r_0)}{\mathcal R(\tau;r_0)}.
        \label{eq:pitch-gradient-Jacobi}
\end{equation}
Thus a zero of $\mathcal R$ is the loss of local invertibility of the map
$r_0\mapsto r(\tau;r_0)$.  If $\mathcal S$ remains nonzero there, the pitch
gradient diverges, but the vanishing of $\mathcal R$ is the fundamental
kinematic criterion.

Differentiating Eq.~\eqref{eq:characteristic-Phi-general} with respect to
$r_0$ gives
\begin{equation}
        \boxed{
        \frac{d\mathcal R}{d\tau}
        =
        -
        \frac{\partial_rK}{\widetilde\Phi^{\,2}}\,
        \mathcal R
        +
        \frac{2K}{\widetilde\Phi^{\,3}}\,
        \mathcal S
        }
        \label{eq:Jacobi-R-equation}
\end{equation}
and
\begin{equation}
        \frac{d\mathcal S}{d\tau}
        =
        \left[
        -\partial_rq
        \left(
        \widetilde\Phi
        +
        \frac{r^2}{\widetilde\Phi}
        \right)
        -
        \frac{2rq}{\widetilde\Phi}
        \right]
        \mathcal R
        -
        q
        \left(
        1-
        \frac{r^2}{\widetilde\Phi^{\,2}}
        \right)
        \mathcal S.
        \label{eq:Jacobi-S-equation}
\end{equation}
All coefficients are evaluated along the characteristic
$r=r(\tau;r_0)$.

Since $\mathcal R(0;r_0)=1$, the exact no-crossing condition on
$0\leq\tau\leq\tau_{\rm f}$ is
\begin{equation}
        \mathcal R(\tau;r_0)>0
        \qquad
        \text{for all }0\leq\tau\leq\tau_{\rm f}
        \text{ and all relevant }r_0.
        \label{eq:no-crossing-condition}
\end{equation}
The first characteristic-caustic time is therefore
\begin{equation}
        \boxed{
        \tau_{\rm c}
        :=
        \inf
        \left\{
        \tau>0:
        \mathcal R(\tau;r_0)=0
        \text{ for at least one }r_0
        \right\}.
        }
        \label{eq:generalized-collapse-time}
\end{equation}
Unlike the time-independent longitudinal-profile case, both the
characteristic speed and the pitch amplitude evolve, so the crossing
criterion cannot in general be reduced to an algebraic condition on the
initial pitch profile alone.

Before the first crossing, Eqs.~\eqref{eq:pitch-gradient-Jacobi} and
\eqref{eq:Jacobi-R-equation} give
\begin{equation}
        \frac{d}{d\tau}\ln\mathcal R
        =
        -\frac{\partial_rK}{\widetilde\Phi^{\,2}}
        +
        \frac{2K}{\widetilde\Phi^{\,3}}
        \partial_r\Phi.
        \label{eq:log-Jacobian-evolution}
\end{equation}
This identity will be used below to connect the characteristic caustic to the
local rotation of the normalized helical field plane.

\subsection{Geometric response of the normalized helical plane}
\label{subsec:geometric-response-general}
The caustic is defined by the characteristic map.  We now ask how the
evolving electromagnetic field responds locally as that map approaches loss
of invertibility.

The normalized rotation rate was defined in
Sec.~\ref{sec:helical-fluxtube}; it is given by Eq.~\eqref{eq:anholonomy-value}
with $\tan\vartheta=\Phi/r$. 
Thus the longitudinal amplitude $B$ drops out explicitly from the normalized
field-plane rotation rate; it enters through the pitch evolution that it
induces.
Writing $\partial_t\Phi=U_\ast\partial_\tau\Phi$ and using
Eq.~\eqref{eq:generalized-pitch-Kq}, one obtains
\begin{equation}
        \boxed{
        \mathcal A_{\rm F}[G]
        =
        U_\ast
        \left[
        \frac{Kr\,\partial_r\Phi}
        {\Phi^2(r^2+\Phi^2)}
        -
        \frac{qr}{\Phi}
        \right].
        }
        \label{eq:anholonomy-master}
\end{equation}
For finite nonzero pitch and regular coefficients, a divergent radial pitch gradient therefore produces divergent field-plane rotation rate.  The lower-pitch
limit $\Phi\to0$ can also produce a singular rotation rate, since
\begin{equation}
        \mathcal A_{\rm F}[G]
        \sim
        U_\ast
        \left[
        \frac{K\,\partial_r\Phi}{r\Phi^2}
        -
        \frac{qr}{\Phi}
        \right],
        \qquad
        \Phi\to0.
        \label{eq:anholonomy-small-pitch}
\end{equation}
In particular, for regular nonzero $q$, the explicit term
$-U_\ast qr/\Phi$ already drives the plane-rotation rate singular.  This limit reappears in the exact homologous solution below, where the pitch slope itself tends to zero while the normalized field-plane rotation rate diverges.

It is useful to distinguish these differential geometric singularities from
divergences of the electromagnetic amplitudes themselves.  The magnitudes of
the two simple sectors are
\begin{equation}
        \mathcal N_G
        =
        \frac{|B|}{r}
        \sqrt{1+\frac{\Phi^2}{r^2}},
        \qquad
        \mathcal N_H
        =
        \frac{|U|}{r}
        \sqrt{1+\frac{r^2}{\Phi^2}}.
        \label{eq:principal-sector-amplitudes}
\end{equation}
If $B$ and $U$ remain finite and nonzero, then
$|\Phi|\to\infty$ drives $\mathcal N_G\to\infty$, whereas
$\Phi\to0$ drives $\mathcal N_H\to\infty$.  These are field-amplitude
singularities, not characteristic caustics.  The two kinds of breakdown may
coexist, but they are conceptually distinct.

The converse limits also require care.  If $U\to0$ together with
$\Phi\to0$, or $B\to0$ together with $|\Phi|\to\infty$, the relevant
sector amplitude can remain finite.  Such a limit may represent a regular
exit from the interior helical branch rather than a field singularity.
Likewise, the vanishing of a sector amplitude is not classified here as a
catastrophe by itself.

\paragraph*{Scalar amplitude rate.}---
For completeness, define
\begin{equation}
        \mathcal C_{\rm F}[G]
        :=
        \partial_t\ln\mathcal N_G.
\end{equation}
The generalized pitch equation gives $ \mathcal C_{\rm F}[G]
        =
        U_\ast K\,\partial_r\Phi/[\Phi(r^2+\Phi^2)].$
Comparison with Eq.~\eqref{eq:anholonomy-master} yields the exact relation
\begin{equation}
        \mathcal A_{\rm F}[G]
        =
        \frac{r}{\Phi}
        \left[
        \mathcal C_{\rm F}[G]-U_\ast q
        \right].
        \label{eq:anholonomy-coherent-relation}
\end{equation}
Hence, as long as the evolution remains in the regular helical interior, with
$0<|\Phi|<\infty$ and regular $r$, $q$, and $U_\ast$,
$ |\mathcal C_{\rm F}|\to\infty
        ~\Longrightarrow~
        |\mathcal A_{\rm F}|\to\infty.$
The scalar amplitude rate therefore does not provide an independent
singularity criterion within the regular finite-pitch regime.  This statement
concerns a differential rate and should not be confused with an actual
field-amplitude divergence
$\mathcal N_G\to\infty$ or $\mathcal N_H\to\infty$.

\paragraph*{Relation to the characteristic Jacobian.}---
Eliminating $\partial_r\Phi$ between
Eqs.~\eqref{eq:log-Jacobian-evolution} and
\eqref{eq:anholonomy-master} yields the exact identity
\begin{equation}
        \boxed{
        \frac{\mathcal A_{\rm F}[G]}{U_\ast}
        =
        \frac{r\widetilde\Phi}
        {2\left(r^2+\widetilde\Phi^{\,2}\right)}
        \frac{d}{d\tau}\ln\mathcal R
        +
        \frac{r\,\partial_rK}
        {2\widetilde\Phi\left(r^2+\widetilde\Phi^{\,2}\right)}
        -
        \frac{qr}{\widetilde\Phi}.
        }
        \label{eq:anholonomy-Jacobian-identity}
\end{equation}
Since $\tan\vartheta=\widetilde\Phi/r$, the coefficient of the logarithmic
Jacobian rate is
\begin{equation}
        \frac{r\widetilde\Phi}
        {2\left(r^2+\widetilde\Phi^{\,2}\right)}
        =
        \frac12\sin\vartheta\cos\vartheta.
        \label{eq:angle-Jacobian-factor}
\end{equation}
This relation makes the distinction between the caustic and its geometric
signature explicit.  The caustic is defined kinematically by
$\mathcal R\to0$, whereas $\mathcal A_{\rm F}$ and
$\mathcal C_{\rm F}$ describe the local response of the electromagnetic field.

Consider a regular helical interior on a compact radial region.  We assume
that there exists a fixed constant \(0<\eta<1\) such that
\begin{equation}
        0<r_-\leq r\leq r_+<\infty,
        \qquad
       \eta
        \leq
        \left|\frac{\widetilde\Phi}{r}\right|
        \leq
        \eta^{-1}  , 
        \label{eq:regular-helical-interior}
\end{equation}
and with $K$, $\partial_rK$, $q$, and $U_\ast$ remaining regular.  
The constant \(\eta\) provides a uniform separation from the limiting configurations \(\widetilde\Phi/r=0\) and
\(|\widetilde\Phi|/r\to\infty\).  Equivalently, the helical angle remains
uniformly away from the boundaries of the regular helical sector.

The
angle $\vartheta$ is then bounded away from both endpoints $0$ and $\pi/2$,
so the coefficient $\sin\vartheta\cos\vartheta$ remains finite and nonzero.
If $\mathcal R\to0$ at finite accumulated time, then
$\ln\mathcal R\to-\infty$, and therefore
$d(\ln\mathcal R)/d\tau$ cannot remain bounded throughout the approach.
Equation~\eqref{eq:anholonomy-Jacobian-identity} then implies that the normalized field-plane rotation rate cannot remain bounded either.

\begin{proposition}[Caustic--field-plane-rotation implication]
\label{prop:general-caustic-anholonomy}
On a compact regular helical interior satisfying
Eq.~\eqref{eq:regular-helical-interior}, assume $K$, $\partial_rK$, $q$, and
$U_\ast$ remain bounded and $U_\ast$ stays bounded away from zero.  If the
characteristic Jacobian reaches $\mathcal R=0$ at finite accumulated time,
then $\mathcal A_{\rm F}[G]$ cannot remain bounded.  For ordinary smooth
first breaking, this unboundedness appears as
$|\mathcal A_{\rm F}|\to\infty$.
\end{proposition}

This result is stronger than the pitch-gradient criterion.  A divergent
$\partial_r\Phi$ is one way to realize the caustic, but it is not required.
If one insists that $\mathcal A_{\rm F}$ remain finite while
$\mathcal R\to0$ under regular background evolution, then the coefficient of
the logarithmic Jacobian rate must itself degenerate.  Equation
\eqref{eq:angle-Jacobian-factor} shows that this requires
\begin{equation}
        \frac{\widetilde\Phi}{r}\to0
        \qquad\text{or}\qquad
        \left|\frac{\widetilde\Phi}{r}\right|\to\infty.
        \label{eq:helical-exit-angle}
\end{equation}
In that limit the normalized helical direction approaches the boundary of
the radial--helical structural class.  Such behavior is therefore a
structural exit from the regular helical interior rather than a caustic of the
regular branch.  Equation~\eqref{eq:helical-exit-angle} is necessary but not
sufficient for a bounded rotation rate, so whether $\mathcal A_{\rm F}$ actually
stays finite in that limit must be decided case by case; the exact homologous
family of Sec.~\ref{subsec:homologous-caustic} is precisely of this type, and
there the rotation rate still diverges.  A structural exit may nonetheless be
nonsingular if the field amplitudes remain finite,
for example through a compensating vanishing of $U$ or $B$.

Field-amplitude blow-up is therefore not a second type of the geometric
caustic studied here.  A divergence of $\mathcal N_G$ or $\mathcal N_H$ may
occur before or together with a characteristic crossing, but it is a distinct
electromagnetic singularity.  Within the regular helical interior, the
intrinsic geometric signature of $\mathcal R=0$ is unbounded, and generically divergent, normalized field-plane rotation rate.

\paragraph*{Physical-time interpretation of the accumulated parameter.}---
If $U_\ast$ vanishes or changes sign, the evolution must be described
piecewise in monotonic $\tau$-patches.  The reference amplitude is related to
the tangential electric field by
\(
U_\ast(t)=r_\ast E_{\hat\phi}(t,r_\ast)
\).
Provided
\begin{equation}
        0<U_{\min}\leq U_\ast(t)\leq U_{\max}<\infty,
        \label{cond1}
\end{equation}
the map $t\mapsto\tau$ is smooth, monotonic, and invertible.  The geometric
caustic does not invalidate this reparametrization.  On every interval
satisfying Eq.~\eqref{cond1}, a caustic at $\tau=\tau_c$ occurs at the
physical time determined by
\begin{equation}
        \int_{t_0}^{t_c}U_\ast(t')\,dt'=\tau_c.
\end{equation}
Thus the accumulated variable changes only the time parametrization of the
caustic as long as the reference electric field remains regular.

The upper bound on $U_\ast$ is not merely a technical condition for the time
reparametrization.  Since
\begin{equation}
        \mathcal N_H^2
        =
        U^2
        \left(
        \frac{1}{r^2}
        +
        \frac{1}{\Phi^2}
        \right),
        \qquad
        U(t,r_\ast)=U_\ast(t),
\end{equation}
one has
\(
\mathcal N_H(t,r_\ast)\geq |U_\ast(t)|/r_\ast
\).
Hence $U_\ast\to\infty$ necessarily produces a divergence of the
complementary-sector field amplitude.  An accompanying divergence of
$\mathcal A_{\rm F}=\partial_t\vartheta
=U_\ast\partial_\tau\vartheta$
would then be driven by singular electromagnetic forcing rather than by the
characteristic geometry itself.  The caustic analysis consequently assumes
Eq.~\eqref{cond1}, thereby separating field-amplitude singularities from the
intrinsic geometric collapse.

\subsection{Combined regularity conditions and benchmark histories}
\label{subsec:regularity-benchmarks}
The preceding analysis separates three notions that should not be identified.
First, characteristic-caustic avoidance requires
\begin{equation}
        \mathcal R(\tau;r_0)>0.
        \label{eq:geometric-regularity-condition}
\end{equation}
Second, electromagnetic field regularity requires the sector amplitudes to
remain bounded from above,
\begin{equation}
        \mathcal N_G<\infty,
        \qquad
        \mathcal N_H<\infty.
        \label{eq:field-amplitude-regularity}
\end{equation}
A vanishing sector amplitude is not classified as a singularity by itself.
Third, persistence within the same nondegenerate radial--helical structural
class requires the helical angle to remain in the interior,
\begin{equation}
        0<\eta
        \leq
        \left|\frac{\Phi}{r}\right|
        \leq
        \eta^{-1}<\infty.
        \label{eq:helical-persistence-condition}
\end{equation}
A fully regular persistent helical branch therefore satisfies all three
conditions.  The first is the geometric criterion of central interest here;
the second excludes an independent field-amplitude blow-up; and failure of the
third may instead represent a regular structural exit from the helical branch.

A simple stabilizing tendency is obtained for a separable longitudinal
profile
\begin{equation}
        B(\tau,r)=T(\tau)g(r),
\end{equation}
for which the function $q$ given just below Eq.~\eqref{eq:generalized-pitch-Kq} becomes
$  q= \dot T/T.$
If $q\leq0$ along a characteristic, Eq.~\eqref{eq:Y-solution} shows that an
initially positive $Y$ cannot reach zero.  If, moreover,
$T(\tau)\to T_\infty>0$ with
\(\int|q|\,d\tau<\infty\), the pitch change remains bounded.  This does not
constitute a global no-caustic theorem: once the relaxation becomes weak, a
residual compressive pitch profile may again enter the time-independent
longitudinal-profile focusing mechanism.  The exact geometric condition
remains $\mathcal R>0$.

\subsection{Exact homologous caustic}
\label{subsec:homologous-caustic}
A power-law ansatz $   B(\tau,r)=b(\tau)r^m$ and
$ \Phi(\tau,r)=\kappa(\tau)r$, with $m\geq0$, admits an exact homologous solution, analyzed in App.~\ref{app:homologous}.  
Here \(m\) and the reference radius
\(r_\ast>0\) are constants, whereas \(b=b(\tau)\) and
\(\kappa=\kappa(\tau)\) evolve with the accumulated time \(\tau\).
A dot denotes \(d/d\tau\).

Substituting the ansatz into
\(
\partial_\tau B=\partial_r\mathcal U
\)
and imposing the normalization
\(
\mathcal U(r_\ast)=1
\)
gives
\begin{equation}
        \dot b
        =
        \frac{m+1}{r_\ast^{m+1}},
        \qquad
        \mathcal U(r)
        =
        \left(\frac{r}{r_\ast}\right)^{m+1}.
        \label{eq:homologous-U}
\end{equation}
Thus both \(\dot b\) and \(\mathcal U(r)\) are independent of \(\tau\), and
$   b(\tau)
        =
        b_0
        +
        (m+1)/r_\ast^{m+1} \times
        (\tau-\tau_0).$
The remaining pitch equation integrates to
\begin{equation}
        \boxed{
        \kappa(\tau)^2
        =
        \frac{D(\tau)}
        {(m+1)b(\tau)^2},
        \qquad
        D(\tau)
        :=
        (m+1)\mathcal C
        -
        m b(\tau)^2 ,
        }
        \label{eq:homologous-U-normalized}
\end{equation}
where \(\mathcal C\) is a constant of integration.  The real branch requires
\(D(\tau)>0\).

Unlike the preceding separable flux history, the normalization leaves no
independent freedom in the time dependence of \(b\): its growth rate is fixed
and positive.  For \(m>0\), this exact family therefore provides a benchmark
in which the homologous evolution is driven toward the finite-\(\tau\)
endpoint \(D=0\), rather than a test of the avoidance freedom available in
the \(q\leq0\) branch.

For $m\neq0$, the characteristic map is homologous,
\begin{equation}
        r(\tau,r_0)
        =
        \mathcal R(\tau)r_0,
        \qquad
        \mathcal R(\tau)
        =
        \left(
        \frac{D(\tau)}{D_0}
        \right)^{1/(2m)}.
        \label{eq:homologous-Jacobian}
\end{equation}
For $m>0$ and $b>0$, $D$ decreases and reaches zero at
\begin{equation}
        \boxed{
        b_c^2
        =
        \frac{m+1}{m}\mathcal C,
        \qquad
        \tau_c-\tau_0
        =
        \frac{r_\ast^{m+1}}{m+1}(b_c-b_0).
        }
        \label{eq:homologous-catastrophe-bound}
\end{equation}
At this endpoint $\mathcal R\to0$ simultaneously for every characteristic
label, while $  \partial_r\Phi
        =
        \kappa
        \longrightarrow0.$
This endpoint is not the interior first crossing of
Sec.~\ref{subsec:flux-rope-collapse-time}: for every $\tau<\tau_c$ the
homologous characteristics remain ordered, and at $\tau_c$ all labels
collapse onto the axis simultaneously.  It is a global loss of invertibility
of the radial map rather than a multivalued characteristic crossing, and
there is therefore no associated Rankine--Hugoniot continuation of the
homologous endpoint.  The Eulerian pitch slope is accordingly not an intrinsic
diagnostic of this global Jacobian collapse.

Since $\tan\vartheta=\kappa$, $ \mathcal A_{\rm F}[G]
        =
        U_\ast(t)\dot\kappa/(1+\kappa^2).$
Near $D=0$, one has\(\kappa\propto\mathcal R^m\) and
\(\dot\kappa\propto\mathcal R^{-m}\).
Provided $U_\ast$ remains finite and bounded away from zero,
\begin{equation}
        \mathcal A_{\rm F}[G]
        \propto
        \mathcal R^{-m},
        \qquad
        m>0.
        \label{eq:homologous-anholonomy-scaling}
\end{equation}
Thus the homologous endpoint is a global characteristic caustic with
divergent normalized field-plane rotation rate, even though the pitch slope tends
to zero.  In terms of the exact identity
\eqref{eq:anholonomy-Jacobian-identity}, the reason is transparent:
$\sin\vartheta\cos\vartheta\sim\kappa\sim\mathcal R^m$, while
$d\ln\mathcal R/d\tau\sim\mathcal R^{-2m}$, leaving the product divergent as
$\mathcal R^{-m}$.  Note that this endpoint has $\widetilde\Phi/r=\kappa\to0$
and therefore lies outside the regular helical interior
\eqref{eq:regular-helical-interior} assumed in
Proposition~\ref{prop:general-caustic-anholonomy}: the degeneracy of
$\sin\vartheta\cos\vartheta$ identified in
Eq.~\eqref{eq:helical-exit-angle} is necessary but not sufficient for a bounded
rotation rate, and here the divergence is established by direct computation
rather than by that proposition.

The field amplitudes provide an independent check that this geometric
singularity need not be an amplitude blow-up.  For the homologous solution,
Eq.~\eqref{eq:principal-sector-amplitudes} gives
\begin{equation}
        \mathcal N_G
        =
        |b|\,r^{m-1}\sqrt{1+\kappa^2},
        \qquad
        \mathcal N_H
        =
        \frac{|U_\ast|\,r^m}{r_\ast^{m+1}}
        \sqrt{1+\kappa^{-2}}.
        \label{eq:homologous-sector-amplitudes}
\end{equation}
Along $r=\mathcal R r_0$ with $\kappa\propto\mathcal R^m$, one finds
\(
\mathcal N_H=O(1)
\)
and
\(
\mathcal N_G\sim\mathcal R^{m-1}
\).
Hence the axis-finite benchmark $m=2$ has
$  \mathcal N_G\to0$, $ \mathcal N_H=O(1)$,
and $  |\mathcal A_{\rm F}[G]|\to\infty.$
The Amp\`ere current supplies a sharper electromagnetic diagnostic.  As shown
explicitly in Appendix~\ref{app:homologous}, along a collapsing characteristic
$r=\mathcal R r_0$ one finds, for $m>1/2$ and regular finite nonzero
$U_\ast$,
\begin{equation}
        J^\chi
        \propto
        \mathcal R^{-2m}
        \propto
        \mathcal A_{\rm F}^{\,2}.
        \label{eq:homologous-current-scaling}
\end{equation}
Thus the shrinking helical sector is not becoming electromagnetically quiet.
For the axis-finite benchmark $m=2$,
\begin{equation}
        \boxed{
        \mathcal N_G\sim\mathcal R,
        \qquad
        \partial_r\Phi=\kappa\sim\mathcal R^2,
        \qquad
        \mathcal A_{\rm F}\sim\mathcal R^{-2},
        \qquad
        J^\chi\sim\mathcal R^{-4}.
        }
        \label{eq:homologous-m2-current-hierarchy}
\end{equation}
The global Jacobian collapse therefore occurs without a field-amplitude
blow-up.  The vanishing of $\mathcal N_G$ does not by itself constitute a
singularity; the collapse is the loss of characteristic invertibility and the
associated singular rotation of the normalized helical field plane.  For
$0<m<1$, by contrast, $\mathcal N_G$ also diverges, so an independent
field-amplitude singularity accompanies the geometric caustic.

\begin{proposition}[Homologous current concentration]
\label{prop:homologous-current}
For the exact homologous branch with $m>0$ and regular finite nonzero
$U_\ast$, $\mathcal A_{\rm F}\propto\mathcal R^{-m}$.  For $m>1/2$ the
longitudinal Amp\`ere current satisfies
$J^\chi\propto\mathcal R^{-2m}\propto\mathcal A_{\rm F}^{2}$ along every
fixed noncentral collapsing characteristic.  Hence the endpoint is a genuine
current-concentration event even when the helical-sector amplitude and the
Eulerian pitch slope vanish.
\end{proposition}

In physical terms, the profile coefficient $b$ grows while the rope contracts
radially and untwists, $\mathcal R\to0$ and $\kappa\to0$, yet the
longitudinal current required to support the restricted radial--helical
organization diverges.

The $m=2$ example is called \emph{axis-finite} rather than ``regular-axis'':
$B_{\hat\chi}=B/r\propto r$ and $B_{\hat\phi}=-A\propto r$ are finite and
vanish at the axis, and the physical current components are finite there at
fixed pre-caustic time, but the radial dependence $|\mathbf x_\perp|$ is not
claimed to be $C^1$ in Cartesian coordinates at $r=0$.  Moreover, because the
power-law fields grow toward large radius, the solution is intended on a
finite-radius interior region requiring matching to an exterior field, not as
a global infinite-radius configuration.  It is therefore used as an $r>0$
local cylindrical benchmark rather than as a theorem of Cartesian axis
smoothness or global finite-energy behavior.  The full derivation is given in
Appendix~\ref{app:homologous}.  Figure~\ref{fig:homologous-scaling}
illustrates the separation between the vanishing pitch slope and helical-sector amplitude and the diverging field-plane rotation rate and current.

\begin{figure}[t]
\centering
\includegraphics[width=\linewidth]{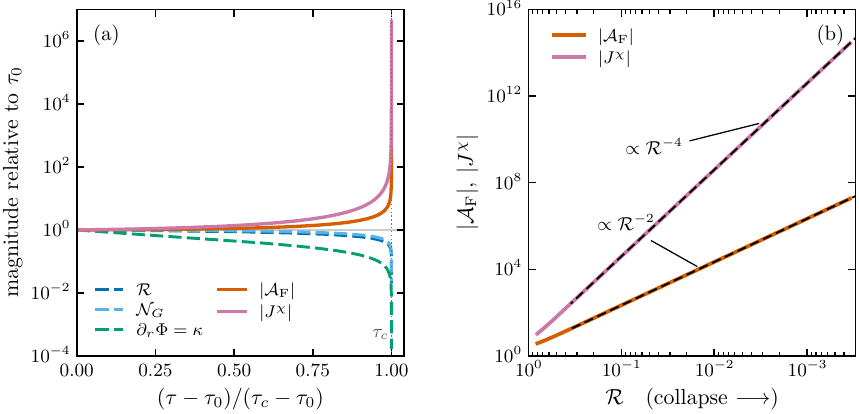}
\caption{Exact $m=2$ homologous benchmark with $U_\ast=r_\ast=\mathcal C=1$ and
$b_{0}=0.5$, evaluated along the characteristic with $r_{0}=1$.
(a) In accumulated time the characteristic Jacobian $\mathcal R$, the
helical-sector amplitude $\mathcal N_{G}$, and the Eulerian pitch slope
$\partial_{r}\Phi=\kappa$ (dashed) all vanish, while the normalized field-plane
rotation $|\mathcal A_{\rm F}|$ and the longitudinal Amp\`ere current (solid)
diverge at the same finite $\tau_{c}$; magnitudes are normalized to their
values at $\tau_{0}$.  Since $\mathcal N_{G}\propto\mathcal R$ for $m=2$,
those two curves coincide.  The range is truncated at
$\mathcal R=2\times10^{-2}$ for legibility.
(b) The same two diverging quantities against $\mathcal R$, which decreases
to the right.  Dashed lines are the predicted asymptotes
$|\mathcal A_{\rm F}|\propto\mathcal R^{-m}=\mathcal R^{-2}$ and
$|J^{\chi}|\propto\mathcal R^{-2m}=\mathcal R^{-4}$ of
Eqs.~\eqref{eq:homologous-anholonomy-scaling} and
\eqref{eq:app-homologous-current-scaling}; logarithmic fits over the
asymptotic interval $10^{-3}\leq\mathcal R\leq10^{-1}$ give slopes $-2.000$
and $-4.000$, confirming Eq.~\eqref{eq:homologous-m2-current-hierarchy}.}
\label{fig:homologous-scaling}
\end{figure}

\section{Matter realization and geometry--matter consistency}
\label{sec:matter-closure}

The preceding sections determine the electromagnetic evolution within the
restricted helical branch, but they do not determine the material system that
supports it.  In particular, the characteristic caustic can be reached while
the electromagnetic field amplitudes and algebraic invariants remain finite,
whereas the Amp\`ere current required by Maxwell's equations generically
becomes singular.  The geometric collapse therefore appears
electromagnetically as a \emph{current-sheet precursor}.  Whether an actual
plasma reaches this endpoint is a separate matter--field consistency problem.

This distinction is essential because the radial pitch characteristics are
not material worldlines determined by ideal flux freezing.  As discussed in
Sec.~\ref{sec:intro}, in the nondegenerate helical branch, there is no timelike four-velocity satisfying $  F_{ab}u^b=0 ,$ and no local frame in which the electric field vanishes entirely.
The electromagnetic characteristics derived in the preceding sections
therefore cannot be identified with a matter flow by ideal-MHD kinematics
alone.  Any alignment between the two must be imposed as an additional
constitutive assumption.

\subsection{Material support as a separate closure problem}
\label{subsec:matter-support}

For a prescribed longitudinal magnetic profile, the geometric construction
determines the pitch evolution, the associated characteristic map, and the
electromagnetic current $J^\mu=\nabla_\nu F^{\nu\mu}$
required to support the field.  It does not, however, provide an evolution law
for the matter variables or for the longitudinal magnetic profile itself.

The persistence of the helical branch is therefore conditional on the
material response.  A constitutive model may alter the longitudinal profile,
change the characteristic geometry, or force the electromagnetic field to
leave the restricted radial--helical class before the formal caustic is
reached.  Conversely, if the material system can continue to supply the
required current while the branch persists, the current concentration found
in Sec.~\ref{sec:helical-fluxtube} follows directly from Maxwell's equations.

A dissipative transport law provides one possible route by which the matter
response can modify the geometric evolution.  Resistive relativistic MHD, for
example, supplies constitutive closures in which electromagnetic gradients
feed back on magnetic-flux transport
\cite{Komissarov2007,PalenzuelaEtAl2009,Andersson2012}.
In a local slender-tube approximation one may schematically write
\begin{equation}
        \partial_\tau B
        \simeq
        D_\tau\,\partial_r^2B,
        \qquad
        D_\tau>0 .
        \label{eq:matter-diffusive-example}
\end{equation}
Such a response can reshape \(B\), and hence the coefficients \(K\) and \(q\)
that govern the characteristic map.  Equation~\eqref{eq:matter-diffusive-example}
is used only as an illustration: the present analysis does not select a
particular resistive law or claim that diffusion is the unique regularizing
mechanism.

\subsection{Aligned matter as a consistency test}
\label{subsec:aligned-matter}

A separate question is whether matter can be forced to follow the geometric
characteristics themselves.  Since this identification does not follow from
the electromagnetic field, it defines an additional constitutive branch.
Appendix~\ref{app:mechanical} tests this assumption for the exact homologous
solution by identifying the material radial motion with the characteristic
map.

The first obstruction is kinematic.  For homologous focusing, the radial map
takes the form~\eqref{eq:homologous-Jacobian}.
If this map is interpreted as a material flow, every noncentral streamline
inherits the corresponding radial velocity.  As the geometric caustic is
approached, this velocity becomes incompatible with relativistic causality
before the formal endpoint is reached.

\begin{proposition}[Causal obstruction of aligned matter]
\label{prop:matter-causal}
If the homologous focusing reaches \(D=0\) at a finite nonzero rate and
\(m>1/2\), every noncentral matter streamline constrained to follow the
characteristic map reaches \(|v_r|=1\) at a time strictly before the
geometric caustic.  The aligned interpretation therefore cannot remain a
causal material flow up to \(\mathcal R=0\).
\end{proposition}

The second obstruction is dynamical.  Even before specifying detailed
dissipative transport, one may ask whether a simple relativistic barotropic
fluid can provide the stress required by the homologous electromagnetic
evolution.  Particle conservation together with
\(
p=w\rho
\)
fixes the scaling of the material density and pressure along the assumed
flow.  Comparison with the radial Euler equation then gives the following
incompatibility.

\begin{proposition}[Linear-barotrope obstruction]
\label{prop:matter-barotrope}
For every nontrivial homologous branch with \(m>1\), \(\beta\neq0\), and a
constant barotropic parameter \(0\leq w\leq1\), relativistic particle
conservation and \(p=w\rho\) are incompatible with the lowest magnetic order
of the radial Euler balance on any open evolving interval.  The detailed
near-axis derivation is given in
Appendix~\ref{app:linear-barotrope}.
\end{proposition}

These two results probe different aspects of the same consistency problem.
The causal obstruction shows that exact characteristic--matter locking cannot
survive to the homologous caustic, whereas the barotropic obstruction shows
that even away from the terminal point a simple perfect-fluid closure does not
generically reproduce the required radial dynamics.  Neither result excludes
more general matter models; nonlinear equations of state, anisotropic
stresses, dissipative transport, or imperfect alignment may change the
conclusion.

A weaker constitutive restriction is considered in
Appendix~\ref{app:physical-scales}.  Imposing a sub-Dreicer thermal bound on
the parallel electric field does not change the geometric threshold
\(\tau_\ast\), but it places a lower bound on the physical time required to
accumulate the drive needed to reach that threshold.

\subsection{Interpretation}
\label{subsec:matter-interpretation}

The preceding results should not be read as saying that the matter must fail
while the electromagnetic geometry continues unchanged.  Rather, they show
that the restricted electromagnetic branch and a chosen material model need
not remain mutually consistent throughout the evolution.

Several outcomes are then possible.  The constitutive response may reshape
the longitudinal magnetic profile and thereby alter the characteristic map;
the matter flow may cease to follow the geometric characteristics; or the
electromagnetic field may reorganize into a more general configuration whose
local field planes no longer belong to the radial--helical class considered
here.  Any of these responses can occur before the formal geometric caustic.

The role of the present construction is therefore diagnostic rather than
constitutive.  It identifies the electromagnetic current and geometric
deformation required to maintain the restricted helical organization.  The
matter dynamics determines whether that organization is physically sustained,
regularized, or abandoned before the ideal current concentration is reached.

\section{Discussion}
\label{sec:discussion}

The central result is a finite-time loss of the restricted radial--helical
organization.  Its defining kinematic feature is the loss of invertibility of
the radial characteristic map.  In the regular helical interior this loss of
invertibility forces the rotation rate of the normalized helical field plane
to become unbounded.  The time-independent longitudinal-profile branch
realizes this through an interior characteristic crossing accompanied by
pitch-gradient steepening.  The exact homologous family provides a distinct
global realization: all radial labels collapse simultaneously toward the
axis while \(\partial_r\Phi\to0\), yet the normalized field-plane rotation
rate diverges.  Pitch-gradient blow-up is therefore only one manifestation of
the geometric collapse, not its defining feature.

The same collapse has a direct electromagnetic signature.  In the
time-independent longitudinal-profile branch, the field amplitudes and
algebraic Maxwell invariants can remain finite while the Amp\`ere current
generically diverges as the characteristic Jacobian vanishes.  We therefore
interpret the formal caustic as a \emph{current-sheet precursor}.  Beyond first
breaking, the reduced convex conservation law admits the standard
entropy-selected Rankine--Hugoniot continuation, which carries a surface
current.  This weak continuation is a well-defined property of the reduced
transport equation, but it does not by itself determine whether a
self-consistent plasma reaches that regime or which finite-width current
layer is selected by resistive, Hall, or kinetic physics.  The homologous
solution gives the complementary global example, with
\(J^\chi\propto\mathcal R^{-2m}\) even though the pitch slope vanishes.

The extensions considered here clarify the robustness and the limits of the
mechanism.  A finite radial magnetic through-flux does not remove the
focusing within the fixed-radial-leg class: in the longitudinally symmetric
sector the characteristic transport is unchanged, while longitudinal
redistribution introduces no intrinsic smoothing scale.  Allowing the
longitudinal magnetic profile itself to evolve changes the characteristic
map and can therefore delay or avoid focusing.  More general deformations of
the local field planes may provide additional regular continuations, but such
solutions leave the restricted radial--helical class and represent a
reorganization of the electromagnetic structure rather than persistence of
the same flux-rope branch.

The physical realization is consequently a matter--field consistency
problem.  The nondegenerate branch satisfies \(\mathbf E\cdot\mathbf B\neq0\), so there is no timelike four-velocity for which \(F_{ab}u^b=0\).  
The radial characteristics are therefore not material trajectories fixed by ideal-MHD flux freezing.  
Constitutive transport may reshape the longitudinal profile,
the matter flow may depart from the geometric characteristics, or the field
may reorganize before the formal caustic is reached.  The causal and
barotropic obstructions of the homologous example illustrate this distinction:
they constrain particular matter realizations without invalidating the
underlying electromagnetic caustic.
Constitutive bounds on the admissible electric field do not change the
geometric threshold itself, but they can place lower bounds on the physical
time required to reach it, as discussed in
Appendix~\ref{app:physical-scales}.

The analysis is local, cylindrical, and restricted to a non-degenerate helical
branch.  It does not provide a self-consistent post-caustic plasma evolution,
nor does it select the resistive, Hall, kinetic, or other microscopic process
that regularizes the current concentration.  Its narrower conclusion is that
a persistent helical electromagnetic organization can possess an intrinsic
nonlinear focusing mechanism: the radial characteristic map can lose
invertibility in finite time, this appears geometrically as an unbounded
rotation rate of the normalized helical field plane, and, under a generic
nondegeneracy condition, the current required to support the branch becomes
singular.  Extending this mechanism to curved spacetime and to
self-consistent kinetic or relativistic plasma closures is a natural next
step.

\section*{Acknowledgment}
This work was supported by the National Research Foundation of Korea (NRF) grants RS-2026-25483539 (H.K.).


%

\vspace{0.2cm}
\noindent


\let\doi\relax



\appendix
\section{Exact homologous family: full derivation}
\label{app:homologous}
This appendix gives the complete derivation of the exact homologous solution
summarized in Sec.~\ref{subsec:homologous-caustic}, together with the special cases \(m=0\) and \(m=2\) and the normalized field-plane rotation rate and Amp\`ere current of the background.  
We restrict to $m\geq0$, which contains all collapsing branches used in the paper.

We work in the normalization \(\partial_\tau B=\partial_r\mathcal U\),
\(\mathcal U(\tau,r_\ast)=1\), and take
\begin{equation}
        B(\tau,r)=b(\tau)r^m,
        \qquad
        \Phi(\tau,r)=\kappa(\tau)r,
        \qquad
        m\geq0.
        \label{eq:app-homologous-ansatz}
\end{equation}
Integrating \(\partial_r\mathcal U=\partial_\tau B=\dot b\,r^m\) with the reference
condition gives
\begin{equation}
        \mathcal U(\tau,r)
        =
        1+\frac{\dot b(\tau)}{m+1}\left(r^{m+1}-r_\ast^{m+1}\right).
        \label{eq:app-homologous-U-general}
\end{equation}
Substituting \eqref{eq:app-homologous-ansatz} into the generalized pitch
equation~\eqref{eq:generalized-pitch-Kq} yields
\begin{equation}
        \dot\kappa
        -
        \frac{\mathcal U}{b\kappa r^{m+1}}
        =
        -\frac{\dot b}{b}\left(\kappa+\frac{1}{\kappa}\right),
        \label{eq:app-homologous-kappa-before}
\end{equation}
and, using \eqref{eq:app-homologous-U-general},
\begin{equation}
        \frac{\mathcal U}{r^{m+1}}
        =
        \frac{\dot b}{m+1}
        +
        \left[1-\frac{\dot b}{m+1}r_\ast^{m+1}\right]\frac{1}{r^{m+1}}.
        \label{eq:app-homologous-U-over-r}
\end{equation}
For \(\kappa\) to depend on \(\tau\) alone, the explicitly \(r\)-dependent term
must vanish, which fixes
\begin{equation}
        \dot b=\frac{m+1}{r_\ast^{m+1}},
        \qquad
        b(\tau)=b_0+\frac{m+1}{r_\ast^{m+1}}(\tau-\tau_0),
        \qquad
        \mathcal U(r)=\left(\frac{r}{r_\ast}\right)^{m+1}.
        \label{eq:app-homologous-U-normalized}
\end{equation}
Equation~\eqref{eq:app-homologous-kappa-before} then reduces to
\begin{equation}
        \dot\kappa
        =
        -\frac{\dot b}{b}\kappa
        -\frac{m}{m+1}\frac{\dot b}{b\kappa},
        \label{eq:app-homologous-kappa-ode}
\end{equation}
which, multiplied by \(\kappa\) and integrated with the factor \(b^2\), gives
\begin{equation}
        \kappa^2(\tau)
        =
        \frac{\mathcal C}{b(\tau)^2}
        -\frac{m}{m+1}
        =
        \frac{D(\tau)}{(m+1)b(\tau)^2},
        \qquad
        D(\tau):=(m+1)\mathcal C-mb(\tau)^2,
        \label{eq:app-kappa-D}
\end{equation}
with \(\mathcal C\) an integration constant; the real homologous branch requires
\(D(\tau)>0\).
The radial characteristic equation becomes
\begin{equation}
        \frac{dr}{d\tau}
        =
        -\frac{r^2\mathcal U}{B\Phi^2}
        =
        -\frac{b\dot b}{D}\,r.
        \label{eq:app-homologous-characteristic}
\end{equation}
Since \(\dot D=-2mb\dot b\), this integrates for \(m\neq0\) to
\begin{equation}
        r(\tau,r_0)=r_0\left[\frac{D(\tau)}{D_0}\right]^{1/(2m)},
        \qquad
        \mathcal R(\tau,r_0)=\left[\frac{D(\tau)}{D_0}\right]^{1/(2m)},
        \qquad
        D_0:=D(\tau_0).
        \label{eq:app-homologous-Jacobian}
\end{equation}
For \(m>0\), \(b>0\), Eq.~\eqref{eq:app-homologous-U-normalized} gives
\(\dot b>0\), hence \(\dot D=-2mb\dot b<0\): the normalized branch focuses.  Its
exact regularity region is
\begin{equation}
        D(\tau)>0
        \quad\Longleftrightarrow\quad
        b(\tau)^2<b_c^2,
        \qquad
        b_c^2:=\frac{m+1}{m}\mathcal C,
        \label{eq:app-homologous-regular-bound}
\end{equation}
and at \(D(\tau_c)=0\) one has \(\kappa^2\to0\) and \(\mathcal R\to0\)
simultaneously, so that amplitude degeneration and characteristic collapse occur at the same threshold,
\begin{equation}
        b(\tau_c)^2=b_c^2=\frac{m+1}{m}\mathcal C,
        \qquad
        \tau_c-\tau_0=\frac{r_\ast^{m+1}}{m+1}\left(b_c-b_0\right)
        \label{eq:app-homologous-catastrophe}
\end{equation}
for initial data \(0<b_0<b_c\).  Within this exact family
\eqref{eq:app-homologous-catastrophe} is both necessary and sufficient for the
map to become singular.

\paragraph{The case \(m=0\).}
The normalization gives \(\dot b=1/r_\ast\), \(\mathcal U=r/r_\ast\),
\(\kappa^2=\mathcal C/b^2\), and \(D=\mathcal C\).  The characteristic Jacobian is
\begin{equation}
        \mathcal R(\tau,r_0)
        =
        \exp\!\left[-\frac{b(\tau)^2-b_0^2}{2\mathcal C}\right]>0
        \label{eq:app-m0-J}
\end{equation}
for every finite \(\tau\): the \(m=0\) branch has no finite-time characteristic
catastrophe, although \(\mathcal R\to0\) asymptotically as \(b\to\infty\).

\paragraph{The axis-finite benchmark \(m=2\).}
Here
\begin{equation}
        B=b(\tau)r^2,
        \qquad
        \dot b=\frac{3}{r_\ast^3},
        \qquad
        \mathcal U=\left(\frac{r}{r_\ast}\right)^3,
        \qquad
        D=3\mathcal C-2b^2,
\end{equation}
so that
\begin{equation}
        \mathcal R(\tau,r_0)
        =
        \left[\frac{3\mathcal C-2b(\tau)^2}{3\mathcal C-2b_0^2}\right]^{1/4},
        \qquad
        b(\tau_c)^2=\frac{3}{2}\mathcal C,
        \qquad
        \tau_c-\tau_0=\frac{r_\ast^3}{3}\left[\sqrt{\tfrac{3\mathcal C}{2}}-b_0\right].
        \label{eq:app-m2}
\end{equation}

\paragraph{Normalized field-plane rotation rate of the homologous background.}
The normalized field-plane rotation rate \(\mathcal A_{\rm F}[G]\) defined in
Eq.~\eqref{eq:anholonomy-def}---the normalized Frobenius scalar of
Sec.~\ref{sec:helical-fluxtube}---is particularly simple on the homologous
ansatz.  Since
\(\Phi=\kappa r\), the unit helical angle in
Eq.~\eqref{eq:anholonomy-value} obeys $
        \tan\vartheta=\kappa(\tau),$
and hence
\begin{equation}
        \mathcal A_{\rm F}[G]
        =
        \partial_t\vartheta
        =
        U_\ast(t)\,
        \frac{\dot\kappa}{1+\kappa^2} .
        \label{eq:app-homologous-normalized-anholonomy}
\end{equation}
Differentiating Eq.~\eqref{eq:app-kappa-D}, or equivalently solving
Eq.~\eqref{eq:app-homologous-kappa-ode} for \(\dot\kappa\), gives
\begin{equation}
        \dot\kappa
        =
        -\frac{\dot b}{b}\kappa
        -\frac{m}{m+1}\,\frac{\dot b}{b\kappa} .
\end{equation}
As \(D\to0\), \(\kappa\to0\) while \(\dot b/b\) remains finite and nonzero,
so $
        \dot\kappa
        \sim
        -\frac{m}{m+1}\,\frac{\dot b}{b_c}\,\frac{1}{\kappa} .$
Since \(\kappa^2=D/[(m+1)b^2]\) and
\(D=D_0\mathcal R^{2m}\), one has
\(\kappa\propto\mathcal R^m\) and therefore
\begin{equation}
        \boxed{
        \mathcal A_{\rm F}[G]
        \;\propto\;
        U_\ast(t)\,\mathcal R^{-m} .
        }
\end{equation}
If \(U_\ast(t)\) remains finite and bounded away from zero, the rotation rate of the helical plane diverges for every \(m>0\), including the
$m=2$ benchmark.  This conclusion is independent of the vanishing
Eulerian pitch slope \(\partial_r\Phi=\kappa\to0\).  The earlier
unnormalized quantity \(\star(e\wedge dG)\) carries an additional factor
\(\mathcal N^2\) and can therefore have a different power law when the sector
amplitude shrinks; only the normalized scalar \(\mathcal A_{\rm F}\) measures
the rotation rate of the field plane itself, and it is the normalized form
that should be used as the geometric diagnostic.

\paragraph{Amp\`ere current of the homologous background.}
The longitudinal current of the evolving longitudinal-profile branch is
\begin{equation}
        J^\chi
        =
        \frac{1}{r}
        \left[
        \partial_t\left(\frac{rU}{\Phi}\right)
        +
        \partial_r(rA)
        \right].
\end{equation}
For the exact homologous solution,
$U=U_\ast(t)(r/r_\ast)^{m+1}$,
$\Phi=\kappa r$, and $A=b\kappa r^{m-1}$, so
\begin{equation}
        \boxed{
        J^\chi
        =
        \frac{r^m}{r_\ast^{m+1}}
        \left[
        \frac{1}{\kappa}\frac{dU_\ast}{dt}
        -
        \frac{U_\ast^2}{\kappa^2}\frac{d\kappa}{d\tau}
        \right]
        +
        mb\kappa r^{m-2}.
        }
        \label{eq:app-homologous-current}
\end{equation}
Along a collapsing characteristic $r=\mathcal R r_0$,
$\kappa\propto\mathcal R^m$ and
$d\kappa/d\tau\propto\mathcal R^{-m}$.  If $U_\ast$ and $dU_\ast/dt$ remain
finite, with $U_\ast$ bounded away from zero, the three terms in
Eq.~\eqref{eq:app-homologous-current} scale respectively as
$O(1)$, $\mathcal R^{-2m}$, and $\mathcal R^{2m-2}$.  Hence for
$m>1/2$ the temporal electric-field term dominates and
\begin{equation}
        \boxed{
        J^\chi
        \propto
        \mathcal R^{-2m}.
        }
        \label{eq:app-homologous-current-scaling}
\end{equation}
Since Eq.~\eqref{eq:app-homologous-normalized-anholonomy} gives
$\mathcal A_{\rm F}\propto\mathcal R^{-m}$, the same limit obeys
\begin{equation}
        \boxed{
        J^\chi
        \propto
        \mathcal A_{\rm F}^{\,2},
        \qquad m>\frac12,
        }
        \label{eq:app-current-anholonomy-square}
\end{equation}
with a finite nonzero proportionality factor on each fixed noncentral
characteristic.  The azimuthal coordinate component is
\begin{equation}
        J^\phi
        =
        -\frac{\dot U_\ast}{r_\ast^{m+1}}r^{m-1}
        +(m-1)b\,r^{m-3},
        \label{eq:app-homologous-current-phi}
\end{equation}
so $J^\phi\sim\mathcal R^{m-3}$ along the collapsing map, whereas the physical
orthonormal component is $J^{\hat\phi}=rJ^\phi\sim\mathcal R^{m-2}$ when the
second term is present.  Thus for $m=2$ the physical azimuthal current remains
finite while the longitudinal current diverges as $\mathcal R^{-4}$.

Because $E_{\hat\chi}=-U/\Phi=O(1)$ on each fixed noncentral homologous
characteristic and $E_{\hat\phi}=U/r=O(\mathcal R^m)$, the electromagnetic
work density is dominated by the longitudinal term,
\begin{equation}
        \mathbf J\cdot\mathbf E
        \sim
        -\frac{U}{\Phi}J^\chi
        \propto
        \mathcal R^{-2m},
        \qquad m>\frac12,
        \label{eq:app-homologous-JE}
\end{equation}
under the same nondegeneracy assumptions.  This is a divergence of field-to-matter
power transfer; interpreting it as irreversible Ohmic heating requires an
additional resistive constitutive law.

For $m=2$ one therefore has $J^\chi\propto\mathcal R^{-4}$ while
$\mathcal N_G\propto\mathcal R$ and $\partial_r\Phi=\kappa\propto\mathcal
R^2$.  The homologous endpoint is a current-concentration event even though
the primary-sector amplitude and Eulerian pitch slope both vanish.

\section{Mechanical response near the homologous catastrophe}
\label{app:mechanical}

This appendix examines whether matter can arrest the exact homologous
collapse of Appendix~\ref{app:homologous}.  Two logically distinct
questions must be separated.  The closure equations of
Sec.~\ref{sec:time-dependent-B} determine the electromagnetic kinematics---the
pitch characteristic map $r=r(\tau;r_{0})$ and its Jacobian
$\mathcal R=\partial r/\partial r_{0}$---but they do \emph{not} determine the matter four-velocity.  
The characteristic map is determined by the electromagnetic evolution and is not, by itself, a material flow map.
Before any statement about density
compression can be made, the relation between the two must be specified.

We therefore proceed in three steps.  First (App.~\ref{app:alignment}) we
make explicit the constitutive assumption under which the geometric
focusing acts as a material compression.  Second
(App.~\ref{app:superluminal}) we show that, in the finite-rate homologous
regime, this aligned flow becomes superluminal before the caustic is
reached, so the aligned picture cannot even be continued up to $\mathcal R=0$ within
a causal medium.  Third (App.~\ref{app:linear-barotrope}) we test a thermodynamically
consistent constant-$w$ linear barotrope and show that particle conservation
fixes a density-amplitude history incompatible with the lowest magnetic order
of the radial Euler equation on any nontrivial evolving homologous branch.

\subsection{Alignment assumption}
\label{app:alignment}
The mechanical response is controlled by how matter flows relative to the
pitch characteristics.  We consider the branch in which the matter
streamlines are aligned with the pitch characteristics.  Recalling that the
accumulated time obeys $d\tau/dt=U_{\ast}(t)$, the characteristic radial
speed in \emph{physical} time is
\begin{equation}
        \frac{u^{r}}{u^{t}}
        =
        \left.\frac{dr}{dt}\right|_{\mathrm{char}}
        =
        U_{\ast}(t)\left.\frac{dr}{d\tau}\right|_{\mathrm{char}}
        =
        -\,U_{\ast}(t)\,\frac{K}{\Phi^{2}}.
        \label{eq:appC-alignment}
\end{equation}
This alignment is a constitutive choice, not a consequence of $dF=0$ or of
ideal flux freezing.  The deeper obstruction is already geometric:
non-degeneracy gives $F\wedge F\neq0$, so the zero-through-flux field
$F$ has no kernel and therefore admits no ideal-MHD four-velocity.  In the static cylindrical frame this appears as
$\mathbf E\parallel\mathbf B$ and $\mathbf E\times\mathbf B=0$, leaving no
nontrivial field-determined drift to identify with the radial characteristics.
Equation~\eqref{eq:appC-alignment} therefore selects the particular matter
response in which the geometric caustic is deliberately realized as a material
compression.  The factor $U_{\ast}(t)$, which was scaled out of the pitch
dynamics, reappears here because it sets the physical-time rate of the aligned
flow.

Under this assumption, and using the exact homologous map of
Appendix~\ref{app:homologous},
\begin{equation}
        r(t,r_{0})
        =
        \mathcal R(t)\,r_{0},
        \qquad
        \mathcal R
        =
        \left(\frac{D}{D_{0}}\right)^{1/(2m)},
        \qquad
        D(\tau)=(m+1)\mathcal C-m\,b(\tau)^{2},
        \label{eq:appC-homologous-map}
\end{equation}
relativistic particle conservation with fixed $\chi$-length fixes the proper
number density along each fluid worldline.  Since
$\nabla_\mu(nu^\mu)=0$, cylindrical symmetry gives
$\partial_t(rnu^t)+\partial_r(rnu^r)=0$, and the homologous map implies
$ n(t,r)\,\gamma(t,r)\,\mathcal R(t)^2
        =
        n_0(r_0)\,\gamma_0(r_0)$ with $r_0={r/\mathcal R(t)} .$
Thus
\begin{equation}
        n(t,r)
        =
        \frac{n_0(r_0)\gamma_0(r_0)}{\gamma(t,r)\mathcal R(t)^2} .
        \label{eq:appC-density-explicit}
\end{equation}
The geometric Jacobian alone therefore does not determine an exact density blow-up: the Lorentz factor of the assumed material flow enters explicitly,
and the aligned flow in fact reaches the luminal obstruction discussed below
before the geometric endpoint.  On every fixed pre-luminal time slice,
however, $\gamma=1+O(r^2)$ and $\gamma_0=1+O(r_0^2)$ near the axis.  This is
sufficient for the leading-order barotropic obstruction derived below.

\subsection{Superluminal flow before the caustic}
\label{app:superluminal}
We assume that $U_*(t)$ remains finite, nonzero, and regular near the terminal time, so that a simple zero of $D(\tau)$ is also a simple zero in physical time. 

Suppose the focusing reserve reaches zero at a finite rate,
\begin{equation}
        D(t)\simeq d_{\ast}\,(t_{c}-t),
        \qquad
        d_{\ast}>0,
        \label{eq:appC-finite-rate}
\end{equation}
so that $\mathcal R(t)\propto(t_{c}-t)^{1/(2m)}$.  The aligned radial velocity of a
fluid element with Lagrangian label $r_{0}$ is
\begin{equation}
        v_{r}(t)
        =
        \dot{\mathcal R}(t)\,r_{0}
        \;\propto\;
        (t_{c}-t)^{\frac{1}{2m}-1}\,r_{0}.
        \label{eq:appC-vr}
\end{equation}
For $m>\tfrac12$ the exponent $\tfrac{1}{2m}-1$ is negative, so $|v_{r}|$
diverges as $t\to t_{c}$.  Any noncentral fluid element with $r_0 \neq 0$ therefore reaches
$|v_{r}|=1$ at a time $t_{v}<t_{c}$, strictly \emph{before} $\mathcal R$ reaches
zero:
\begin{equation}
        m>\tfrac12
        \quad\Longrightarrow\quad
        |v_{r}(t_{v})|=1
        \ \text{ at some }\ t_{v}<t_{c}.
        \label{eq:appC-tv}
\end{equation}
The immediate consequence is that the aligned homologous flow cannot be
continued as a causal matter motion all the way to the geometric caustic.
The obstruction here is not the stiffness of the pressure but the kinematics
of the assumed flow itself: at $t_{v}$ the self-similar velocity field
saturates the light cone.  The mechanical question of pressure support is
therefore meaningful only on the pre-luminal interval $t<t_{v}$, and it is
on that interval that the following equation-of-state analysis applies.

\subsection{Overdetermination for a linear barotrope}
\label{app:linear-barotrope}

We next test whether the aligned homologous geometry can satisfy the radial
momentum balance of a thermodynamically consistent linear barotrope.  We show
that it cannot: the obstruction already appears at the lowest spatial magnetic
order and is therefore independent of the Lorentz-factor corrections, which
enter only at higher near-axis radial order, and of the physical-time
normalization $U_{\ast}(t)$.  We restrict the claim to a linear barotrope
$p=w\rho$ with $w$ constant; a general nonlinear $p(\rho)$ can source
additional radial powers and is left open.

We use the accumulated variable $\tau$ only to express the known homologous
coefficients, in which $b(\tau)=b_{0}+\beta\tau$ is linear and
$D(\tau)=(m+1)\mathcal C-mb^{2}$.  Because the no-go result below rests on the spatial magnetic order alone, the
velocity- and acceleration-dependent terms at the higher radial order
$r^{\,2m-1}$---including the higher-order Lorentz-factor corrections---are not
needed and are not used.

Evaluated on the homologous fields $B=b\,r^{m}$, $A=b\kappa\,r^{m-1}$,
$\Phi=\kappa r$, $U=(r/r_{\ast})^{m+1}$, the radial Lorentz force density
separates into two radial powers,
\begin{equation}
        f_{r}
        =
        c_{\mathrm{mag}}(\tau)\,r^{\,2m-3}
        +
        c_{\mathrm{comp}}(\tau)\,r^{\,2m-1},
        \qquad
        c_{\mathrm{mag}}
        =
        \frac{mD}{m+1}+(m-1)b^{2} .
        \label{eq:appC-fr-split}
\end{equation}

\paragraph{Density scaling of a linear barotrope.}
For $p=w\rho$ with $w$ constant, thermodynamic consistency
$d\rho=\bigl[(\rho+p)/n\bigr]\,dn$ \cite{Anile1989,RezzollaZanotti2013} integrates to
$ \rho\propto n^{\,1+w}.$
From Eq.~\eqref{eq:appC-density-explicit}, a power-law core
$n_0(r_0)=n_{00}r_0^s[1+O(r_0^2)]$ has, on each fixed pre-luminal time slice,
\begin{equation}
        n(t,r)
        =
        n_{00}\mathcal R^{-(s+2)}r^s\left[1+O(r^2)\right].
\end{equation}
The Lorentz-factor correction therefore enters two radial orders above the
leading core term.  Thermodynamic consistency then gives
\begin{equation}
        \rho
        =
        \rho_{00}\mathcal R^{-(s+2)(1+w)}r^{\,s(1+w)}
        \left[1+O(r^2)\right].
        \label{eq:appC-rho-scaling}
\end{equation}
Matching the leading radial pressure-gradient power $r^{s(1+w)-1}$ to the magnetic Lorentz-force power $r^{2m-3}$ fixes the core index through $  s(1+w)=2m-2,$
this matching is not an extra assumption but the only way to balance the
$r^{2m-3}$ order at all: no other term in the radial momentum balance carries
that power, so if $s(1+w)-1\neq 2m-3$ the magnetic-order forcing is already
unbalanced and the obstruction follows immediately.  With the exponents
matched, the leading near-axis density coefficient carries the definite time dependence
\begin{equation}
        \rho
        =
        \varrho(\tau)\,r^{\,2m-2}\left[1+O(r^2)\right],
        \qquad
        \varrho(\tau)
        =
        \varrho_{0}\,\mathcal R^{-(2m+2w)}
        =
        \varrho_{0}\left(\frac{D_{0}}{D}\right)^{\!(m+w)/m},
        \label{eq:appC-rho-amp}
\end{equation}
where $\varrho_{0}$ is set once by the initial profile and $D_{0}=D(\tau_{0})$.
\paragraph{When $m=1$.}
For $m\geq1$ and $D>0$, $c_{\mathrm{mag}}>0$.  In particular, for $m=1$
the pressure gradient vanishes whereas $c_{\mathrm{mag}}=D/2>0$, so the
magnetic-order balance is impossible. We therefore assume $m>1$ below.

\paragraph{The magnetic-order balance is overdetermined.}
At the purely spatial magnetic order, the radial pressure gradient is
$\partial_{r}p=(2m-2)\,w\varrho\,r^{2m-3}$, and the radial momentum balance
requires
\begin{equation}
        (2m-2)\,w\,\varrho(\tau)
        =
        c_{\mathrm{mag}}(\tau).
        \label{eq:appC-mag-order}
\end{equation}
Inserting \eqref{eq:appC-rho-amp} gives
\begin{equation}
        w\,\varrho_{0}
        =
        \frac{c_{\mathrm{mag}}(\tau)}{2(m-1)}
        \left(\frac{D(\tau)}{D_{0}}\right)^{\!(m+w)/m}.
        \label{eq:appC-wrho0}
\end{equation}
The left-hand side is a constant.  The right-hand side depends on $\tau$
through $b(\tau)$ and $D(\tau)$, and its logarithmic derivative
\begin{equation}
        \frac{d}{d\tau}\ln
        \!\left[
        c_{\mathrm{mag}}\,D^{(m+w)/m}
        \right]
        =
        \frac{c_{\mathrm{mag}}'}{c_{\mathrm{mag}}}
        +
        \frac{m+w}{m}\,\frac{D'}{D}
        \label{eq:appC-logderiv}
\end{equation}
does not vanish identically for generic $m$ and $w$: with
$c_{\mathrm{mag}}'=-2\beta b/(m+1)$ and $D'=-2m\beta b$,
Eq.~\eqref{eq:appC-logderiv} is a nonzero rational function of
$b(\tau)$.  Hence no constant pair $(w,\varrho_{0})$ satisfies
\eqref{eq:appC-wrho0}: the aligned magnetic-order balance has no
linear-barotropic solution, before the higher-order $r^{\,2m-1}$ balance is
even imposed.
For Eq.~\eqref{eq:appC-logderiv} to vanish identically, the polynomial identity
\begin{equation}
        D+(m+w)(m+1)c_{\rm mag}
        =(m+1)\mathcal C[1+m(m+w)]-(2m+w)b^2
        =0
        \label{eq:appC-polynomial-identity}
\end{equation}
must hold for all $b$.  Its two coefficients require $m=1$ and $w=-2$,
which is outside the physical range and is already excluded by the $m=1$
magnetic-order balance above.  Therefore, for every nontrivial homologous
branch with $m>1$, $\beta\neq0$, and $0\leq w\leq1$, the logarithmic
derivative in Eq.~\eqref{eq:appC-logderiv} cannot vanish identically on any
open time interval.

\paragraph{Statement of the obstruction.}
On the exact homologous branch the aligned momentum balance admits no
linear-barotropic perfect-fluid support.  With the density fixed by particle
conservation and by the consistency relation $\rho\propto n^{1+w}$, the
single constant $w$ (together with the fixed amplitude $\varrho_{0}$) cannot
match the determined time dependence of the magnetic-order forcing.  The
argument uses only the homologous scalings, particle conservation, and the
spatial radial gradient of the pressure; it is independent of $U_{\ast}(t)$ and, at the leading near-axis order used here, of the Lorentz factor $\gamma$.

\paragraph{Scope.}
The result is restricted to a linear equation of state.  A nonlinear
contribution of the form $\lambda\,\rho^{1+2/q}$ with $q=2m-2$, when evaluated on a density component scaling as $r^q$, can produce a pressure gradient at the higher radial order $r^{\,2m-1}$. 
So it is not excluded by the argument above.  Whether such nonlinear or anisotropic matter can
consistently support the aligned homologous focusing is left open here.

\section{Dreicer-limited physical scale of the time-independent longitudinal-profile branch caustic}
\label{app:physical-scales}

The fixed-flux threshold derived in Sec.~\ref{sec:helical-fluxtube} is expressed naturally in terms of the accumulated drive
\(
\tau=\int U\,dt
\).
Here we translate that threshold into directly measurable electromagnetic quantities and ask how rapidly it can be reached while the accompanying parallel electric field remains within a collisional, sub-Dreicer thermal regime.  SI units are restored only in this appendix.

In the orthonormal frame of the static cylindrical observers,
\begin{equation}
        B_{\hat\chi}=\frac{B}{r},
        \qquad
        B_{\hat\phi}=-A,
        \qquad
        \tan\vartheta
        =
        \frac{|B_{\hat\phi}|}{B_{\hat\chi}}
        =
        \frac{|\Phi|}{r}.
        \label{eq:appD-observable-pitch}
\end{equation}
Hence $  |\Phi|
        =r|B_{\hat\phi}|/B_{\hat\chi} .$
In the magnetoquasistatic limit, where the displacement-current
contribution to Amp\`ere's law is negligible,
\begin{equation}
        |B_{\hat\phi}|
        \simeq
        \frac{\mu_0 |I_{\rm enc}(r)|}{2\pi r} \quad \Rightarrow \quad
        |\Phi|
        \simeq
        \frac{\mu_0 |I_{\rm enc}(r)|}
             {2\pi B_{\hat\chi}(r)} .
        \label{eq:appD-pitch-current}
\end{equation}
For a positive-pitch branch the compressive condition
\(
\partial_r\Phi<0
\)
can therefore be interpreted, within this quasistatic approximation, as an
outward decrease of the enclosed axial current relative to the longitudinal
field.  A return-current layer provides one possible realization.

To convert the geometric threshold to physical time, write
\begin{equation}
        \partial_r\Phi
        =
        -\frac{\Phi}{L_c},
        \qquad
        L_c>0,
        \label{eq:appD-compressive-scale}
\end{equation}
where \(L_c\) is the local radial scale of the compressive region.
Equation~\eqref{eq:local-collapse-time} then gives
\begin{equation}
        \tau_\ast
        =
        \frac{B L_c\Phi^2}{2r^2}.
        \label{eq:appD-tau-scale}
\end{equation}
If $U$ varies slowly on the focusing time scale,
so that \(\tau_\ast\simeq |U|t_\ast\), the orthonormal electric
components
\begin{equation}
        E_{\hat\phi}=\frac{U}{r},
        \qquad
        E_{\hat\chi}=-\frac{U}{\Phi}
\end{equation}
give
\begin{equation}
        t_\ast
        \simeq
        \frac{L_c B_{\hat\chi}\tan^2\vartheta}
             {2|E_{\hat\phi}|}
        =
        \frac{L_c B_{\hat\chi}\tan\vartheta}
             {2|E_{\hat\chi}|}
        =
        \frac{L_c}{2|v_{\rm c}|},
        \label{eq:appD-physical-collapse-time}
\end{equation}
where
\begin{equation}
        |v_{\rm c}|
        =
        \frac{|E_{\hat\chi}|}
             {B_{\hat\chi}|\tan\vartheta|}
        =
        \frac{|E_{\hat\phi}|}
             {B_{\hat\chi}\tan^2\vartheta}
        \label{eq:appD-characteristic-speed}
\end{equation}
is the local radial characteristic speed in physical time.
For a general time-dependent drive the exact statement remains
$  \int_{t_0}^{t_\ast}U(t')\,dt'
        =
        \tau_\ast .$

\paragraph*{Dreicer-limited thermal branch.}---
The zero-through-flux branch satisfies
\(\mathbf E\parallel\mathbf B\).  Hence the electric field entering the
runaway criterion is the full parallel magnitude,
\begin{equation}
        E_\parallel
        =
        |\mathbf E|
        =
        |E_{\hat\chi}|
        \sqrt{1+\tan^2\vartheta}.
        \label{eq:appD-Eparallel}
\end{equation}
For a fully ionized collisional plasma, the conventional Dreicer field is
\begin{equation}
        E_D
        =
        \frac{
        e^3 n_e \ln\Lambda
        }{
        4\pi\epsilon_0^2 k_B T_e
        },
        \label{eq:appD-Dreicer}
\end{equation}
where \(n_e\), \(T_e\), and \(\ln\Lambda\) are the electron density,
temperature, and Coulomb logarithm, respectively
\cite{Dreicer1959}.

We use \(E_D\) not as an absolute upper bound on electric fields in a
collisionless or runaway plasma, but as the boundary of the
\emph{sub-Dreicer collisional thermal branch}.  Requiring
$    E_\parallel\leq E_D$
implies
\begin{equation}
        |E_{\hat\chi}| \leq  |\cos\vartheta| E_D,
\end{equation}
and therefore bounds the characteristic speed from above,
\begin{equation}
        \boxed{
        |v_{\rm c}|
        \leq
        v_{{\rm c},D}
        :=
        \frac{E_D}{B_{\hat\chi} } \left| \frac{\cos \vartheta}{\tan\vartheta} \right| .
        }
        \label{eq:appD-Dreicer-vc}
\end{equation}
The same condition gives a lower bound on the physical focusing time,
\begin{equation}
        \boxed{
        t_\ast
        \geq
        t_{\ast,D}
        :=
        \frac{
        L_c B_{\hat\chi}
        }{
        2E_D
        }  \left| \frac{\tan \vartheta}{\cos\vartheta} \right|.
        }
        \label{eq:appD-Dreicer-time}
\end{equation}
Thus, once the thermal state and the magnetic geometry are specified,
the fastest approach to the geometric threshold within this
sub-Dreicer branch is no longer a freely chosen drive calibration.

Table~\ref{tab:physical-scales} illustrates the resulting bounds.
The coronal entries use
\(n_e=10^{15}\,{\rm m^{-3}}\),
\(T_e=10^6\,{\rm K}\),
and \(\ln\Lambda=20\), giving
\(E_D\simeq6.0\times10^{-3}\,{\rm V\,m^{-1}}\).
The prominence-like benchmark uses
\(n_e=10^{17}\,{\rm m^{-3}}\),
\(T_e=10^4\,{\rm K}\), and
\(\ln\Lambda\simeq10\), while the tokamak benchmark uses
\(n_e=10^{20}\,{\rm m^{-3}}\),
\(T_e=5\,{\rm keV}\), and
\(\ln\Lambda\simeq17\).

\begin{table}[t]
\centering
\caption{
Dreicer-limited characteristic speeds and minimum focusing times.
The bounds apply only to a collisional sub-Dreicer thermal realization,
\(E_\parallel\leq E_D\).
The prominence-like row uses the fully ionized Coulomb estimate only as
a formal benchmark; partial ionization can modify the effective runaway
threshold.
}
\label{tab:physical-scales}

\begin{tabular}{lccc}
\hline\hline
configuration
&
\(E_D\,[{\rm V\,m^{-1}}]\)
&
\(v_{{\rm c},D}\)
&
\(t_{\ast,D}\)
\\
\hline

AR corona,
\(10\,{\rm G}\),
\(L_c=3\,{\rm Mm}\)
&
\(6.0\times10^{-3}\)
&
\(4.2\,{\rm m\,s^{-1}}\)
&
\(3.5\times10^{5}\,{\rm s}
 \simeq4.1\,{\rm d}\)
\\

AR corona,
\(100\,{\rm G}\),
\(L_c=3\,{\rm Mm}\)
&
\(6.0\times10^{-3}\)
&
\(4.2\times10^{-1}\,{\rm m\,s^{-1}}\)
&
\(3.5\times10^{6}\,{\rm s}
 \simeq41\,{\rm d}\)
\\

prominence-like,
\(100\,{\rm G}\),
\(L_c=1\,{\rm Mm}\)
&
\(3.0\times10^{1}\)
&
\(2.1\times10^{3}\,{\rm m\,s^{-1}}\)
&
\(2.4\times10^{2}\,{\rm s}
 \simeq3.9\,{\rm min}\)
\\

tokamak core,
\(2.5\,{\rm T}\),
\(L_c=0.1\,{\rm m}\)
&
\(8.9\)
&
\(2.2\times10^{1}\,{\rm m\,s^{-1}}\)
&
\(2.3\times10^{-3}\,{\rm s}\)
\\
\hline\hline
\end{tabular}
\end{table}

For the first three rows we have taken
\(\tan\vartheta=1\), whereas the tokamak entry uses
\(\tan\vartheta=0.16\).
The estimates display a strong hierarchy.  A hot, rarefied active-region
corona restricted to a sub-Dreicer thermal branch accumulates the geometric
drive only on day-to-week time scales, whereas the same bound allows
minute-scale focusing in a cooler and denser prominence-like plasma and
millisecond-scale focusing for the tokamak parameters shown here.
These are lower bounds within the stated thermal branch, not predictions
of the actual collapse time.

The Dreicer restriction also leaves the branch deeply magnetically
dominated.  Defining
\begin{equation}
        \epsilon_\parallel
        :=
        \frac{|\mathbf E|}{c|\mathbf B|},
        \label{eq:appD-magnetic-dominance}
\end{equation}
the sub-Dreicer condition gives
\begin{equation}
        \epsilon_\parallel
        \leq
        \frac{E_D |\cos \vartheta|}
        {
        cB_{\hat\chi}
        } .
        \label{eq:appD-Dreicer-epsilon}
\end{equation}
Small \(\epsilon_\parallel\) does not restore ideal MHD.  The
nondegenerate helical branch still satisfies
\(G\wedge H\neq0\) and hence \(F\wedge F\neq0\) whenever the drive is
nonzero.  Thus
\begin{equation}
        \boxed{
        \epsilon_\parallel\ll1
        \quad\text{is compatible with}\quad
        F\wedge F\neq0 .
        }
        \label{eq:appD-small-nonideal-amplitude}
\end{equation}

The main implication is therefore not that the geometric caustic must
occur on the time scales listed in Table~\ref{tab:physical-scales}.
Rather, if the helical branch is required to remain a
collisional sub-Dreicer thermal plasma, it cannot reach the fixed-flux
geometric threshold faster than \(t_{\ast,D}\).  Faster focusing requires
a super-Dreicer, collisionless, or otherwise nonthermal mechanism capable
of sustaining the required parallel electric field.  
The geometric threshold itself is unchanged: an arbitrarily weak but sufficiently persistent drive can still accumulate the finite value
\(\tau_\ast\), while the constitutive plasma regime determines how rapidly that threshold can be approached.

\paragraph*{Collisionless contrast: blackhole jets.}---
Relativistic black-hole jets provide a complementary example lying outside the thermal branch represented in Table~\ref{tab:physical-scales}. Near the jet base the plasma is expected to be collisionless and may be supplied by pair creation in regions with unscreened electric field, so the classical Dreicer field is not an appropriate constitutive upper bound on $E_\parallel$.
 In such systems the geometric threshold derived here remains applicable at the electromagnetic level, but its physical-time realization must instead be determined by a kinetic or pair-production closure.

\end{document}